\documentclass[aps,pra,twocolumn,superscriptaddress,10pt]{revtex4-2}

\usepackage[english]{babel}
\usepackage[utf8]{inputenc}

\usepackage{amsthm}
\usepackage{amsmath,bm,bbm}
\usepackage{amssymb}
\usepackage{amsfonts}
\usepackage{graphicx}
\graphicspath{{manuscript_figures/}}
\graphicspath{{./figures/}}
\usepackage{txfonts} 
\usepackage{xcolor}
\usepackage{float}
\usepackage{braket}
\usepackage{booktabs}
\usepackage{comment}

\usepackage{empheq}

\usepackage{makecell}
\usepackage{multirow}
\usepackage[squaren]{SIunits}

\usepackage[colorlinks=true,linkcolor=blue,citecolor=blue,urlcolor=blue]{hyperref}

\usepackage{bbold}

\newcommand{\avg}[1]{\langle#1\rangle}		

\newcommand{\id}{\mathbb{1}} 

\allowdisplaybreaks

\begin{document}

\title{Non-interferometric tests of gravitational decoherence models with mechanical oscillators}

\author{Sandro Donadi{$^\star$}}
\email{sandro.donadi@units.it}
\affiliation{Department of Physics, University of Trieste, 34151 Trieste, Italy.}
\affiliation{Istituto Nazionale di Fisica Nucleare, Trieste Section, 34127 Trieste, Italy.}

\author{Iarley P. Lobo$^\star$}
\email{iarley\_lobo@fisica.ufpb.br}
\affiliation{Physics Department, Federal University of Paraíba, Caixa Postal 5008, 58051--970, João Pessoa, Paraíba, Brazil.}
\affiliation{Physics Department, Federal University of Campina Grande, Caixa Postal 10071, 58429-900 Campina Grande, Paraíba, Brazil.}

\author{Matteo Fadel}
\email{matteofadel@pku.edu.cn}
\affiliation{Institute of Quantum Science and Technology $\&$ Frontiers Science Center for Nano-Optoelectronics, School of Physics, Peking University, Beijing 100871, China}
\affiliation{Department of Physics, ETH Z\"{u}rich, Z\"{u}rich 8093, Switzerland}

\makeatletter
\g@addto@macro\frontmatter@footnote@produce{%
  \begingroup
  \def\@thefnmark{}%
  \frontmatter@footnotetext{%
    $\star$ These authors contributed equally to this work.%
  }%
  \endgroup
}
\makeatother

\begin{abstract}
Phenomenological models of quantum-gravity-induced decoherence can leave observable effects even in experiments that do not rely on interferometric measurements. 
We study three such models, based respectively on deformed relativistic symmetries, stochastic metric fluctuations, and a stochastic generalized uncertainty principle, all of which induce decoherence in the momentum basis. 
For a free particle, the models preserve momentum distribution and mean energy but produce position diffusion and momentum-space decoherence with model-specific momentum dependences. 
For a harmonic oscillator, the noncommutativity of the collapse operators with the oscillator Hamiltonian results in anomalous heating, with the three models leading to qualitatively different energy dynamics. 
The first gives heating at a constant rate, the second at a rate proportional to the kinetic energy, which can lead to exponential growth, and the third at a rate growing cubically with excitation.
We further derive the stationary position power spectral density and show that all three models produce the same leading-order excess lineshape in the weak-noise regime, distinct from that of force noise.
Comparing with existing mechanical platforms, we find that displacement-noise measurements can already constrain the fluctuation time of the first model below the Planck time, while occupation measurements could improve the bounds on the second model by up to fourteen orders of magnitude over those from ground-state heating of levitated particles.
Summing the heating over all phonon modes of a solid, the residual heat release of low-temperature experiments constrains the first model far below the Planck time and the second at the Planck time.
Our results establish heating and continuous measurements as complementary non-interferometric probes of gravity-related decoherence.
\end{abstract}

\maketitle


\section{Introduction}

Reconciling quantum mechanics with general relativity remains one of the central open problems in fundamental physics. In the absence of a complete theory of quantum gravity, a fruitful strategy has been to ask whether low-energy signatures of Planck-scale physics could already be within reach of table-top experiments \cite{Amelino-Camelia:2008aez}. 
A common prediction across a broad class of such phenomenological approaches is that quantum gravity might not only introduce a small, coherent correction to the Schrödinger equation \cite{Kiefer:1990pt,Kiefer:1992,Bosso:2023aht,Das:2008kaa,Wagner:2023fmb,Lemos:2026spy}, but might also give rise to fundamental decoherence channels, since fluctuations at the Planck scale (whether of the spacetime metric, of the canonical commutation relations, or of the light cone itself) can affect a quantum system by driving dissipative dynamics \cite{Blencowe:2012mp,bassi2017gravitational,anastopoulos2022gravitational}. Because this decoherence is universal and may exhibit different scaling laws depending on its microscopic origin, it offers a unique opportunity for exploring low-energy quantum gravity phenomenology with table-top experiments involving massive mechanical quantum systems.

\begin{figure}[h!]
    \centering
    \includegraphics[width=\columnwidth]{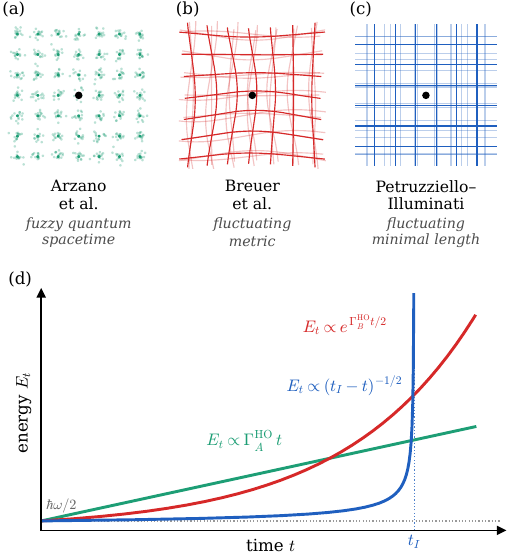}
    \caption{\textbf{Illustrations of the studied models.} (a) Arzano \textit{et al.} deformed relativistic symmetries on a non-commutative spacetime \cite{Arzano:2022nlo}. (b) Breuer \textit{et al.} stochastic metric fluctuations \cite{Breuer:2008rh}. (c) Petruzziello--Illuminati stochastic minimal length \cite{Petruzziello:2020wkd}. In each panel, faint copies of the background pictorially show different noise realizations around a test particle (black dot).
    (d) Qualitative energy growth of an undamped oscillator starting in the ground state: linear for model (a), exponential for model (b), and divergence around $t_I \simeq 0.3/\Gamma_I^{\rm HO}$ for the secular approximation of model (c). The rates $\Gamma^{\rm HO}_{X}$ are given in Eq.~\eqref{eq:HOdecoherenceRates}.}
    \label{fig1}
\end{figure}

Several master equations have been proposed to capture this idea, each rooted in a distinct microscopic picture of Planck-scale physics. We focus here on three models. 
In the first, the relativity principle is deformed on a non-commutative spacetime so as to preserve an invariant energy scale. This results in a momentum-dependent collapse operator that quantifies the residual imprint of the deformed symmetry on non-relativistic dynamics~\cite{Arzano:2022nlo}. 
The second model treats the metric tensor as a stochastic variable fluctuating around its classical background. This leads to a collapse operator proportional to the kinetic energy \cite{Breuer:2008rh}, which, for a free particle, yields the same energy-basis decoherence rate as a thermal graviton bath~\cite{Blencowe:2012mp}.
The third promotes the minimal length of a generalized uncertainty principle to a stochastically fluctuating quantity, which leads to a collapse operator quartic in momentum~\cite{Petruzziello:2020wkd}.
Despite their common Lindbladian structure, these three models encode qualitatively different microscopic physics, and, as we show below, translate into decoherence and heating rates with markedly different dependence on momentum, mass, and oscillation frequency. This raises the prospect of not only bounding, but even discriminating between quantum-gravity models using tailored experiments.

Mechanical oscillators are natural candidates for such tests. Advances in opto- and electromechanics have made it possible to cool motional degrees of freedom of massive particles to their ground state~\cite{Delic2020,Tebbenjohanns2021,Magrini2021,Piotrowski2023}, to prepare and read out nonclassical states of motion~\cite{Bild2023,rahman_genuine_2025}, and to monitor position and momentum with a precision limited only by quantum back-action~\cite{Magrini2021,Tebbenjohanns2021}; see Refs.~\cite{AspelmeyerRMP2014,MillenReview2020} for reviews. These platforms have already been used to place some of the tightest existing bounds on collapse-model and gravity-induced decoherence parameters~\cite{Carlesso2016,Helou2017,Vinante2017,Carlesso2022,Schrinski2022,donadi_quantum_2025}. They have also been proposed as building blocks for more ambitious tests of the quantum nature of gravity itself, for instance via gravitationally induced entanglement between masses.
Unlike a free particle, a harmonic oscillator offers a particularly powerful probe. Since its Hamiltonian generally does not commute with the collapse operators appearing in these models, an oscillator acquires nontrivial dynamics, revealing decoherence and heating signatures that have no free-particle analog and that depend sensitively on its mass and frequency.

In this work we present a systematic, comparative investigation of the three gravity-induced decoherence models just mentioned \cite{Arzano:2022nlo,Breuer:2008rh,Petruzziello:2020wkd} for both a free particle and a harmonic oscillator. For the free particle we summarize closed-form expressions for the momentum-basis decoherence rates and show that they scale as different powers of the momentum spread and mean momentum, providing a first, simple discriminant between models. 
For the harmonic oscillator, we compute the time evolution of the mean phonon number and obtain closed-form expressions for the anomalous heating rates. For the first two models we also derive the stationary position power spectral density, identifying the excess noise that the gravity-induced dissipator adds to the standard spectrum of the damped oscillator. 
Taken together, these results provide concrete, experimentally accessible observables with which current and near-future mechanical oscillator experiments can constrain, and potentially distinguish between, competing scenarios for Planck-scale-induced decoherence.

\section{Models for gravity-related decoherence}

Different low-energy phenomenological approaches to quantum gravity lead to distinct effective master equations, whose predictions can be confronted with experiments. We present the three cases of our interest here below. 
We denote the kinetic energy by $\hat{K}=\hat{p}^2/2m$, with $\hat{p}$ the momentum operator and $m$ the particle mass.
The symbol $\tau_X$ indicates a parameter with dimensions of time that sets the strength of the gravity-related modification in the model labeled by $X=A,B,I$. 
A natural reference scale for this time could be the Planck time $t_P\simeq 5.4\times 10^{-44}\,\mathrm{s}$, although the precise value of $\tau_X$ is model dependent.

The first model we consider is the one by Arzano \textit{et al.}~\cite{Arzano:2022nlo}, which takes the form
\begin{equation}\label{eq:ArzanoModel}
\partial_{t}\hat{\rho}(t)=-\frac{i}{\hbar}\left[\hat{H}_{0},\hat{\rho}(t)\right] - \tau_A \dfrac{c^2}{2\hbar^2} \left[\hat{p},\left[\hat{p},\hat{\rho}(t)\right]\right] \;.
\end{equation}
This model is a consequence of extending the relativity principle to a non-commutative fuzzy spacetime, see Fig.~\ref{fig1}a, whose symmetries are described by a Hopf algebra that deforms the Poincar\'e algebra. One of the main aspects of this approach is a modified composition rule for momenta that, at the quantum level, deforms the way commutators act on operators, with the quantum gravity scale measuring the departure from special relativity. 
In this model, it is possible to have an invariant energy scale in a relativistic setting by deforming the Lorentz transformations and the energy-momentum conservation laws. 
At the non-relativistic level, the model still carries terms that depend on the quantum gravity scale. For this reason, when the deformed algebra is used to describe a quantum system, the evolution of the density matrix gains an additional term that has a Lindblad dissipator term as given in Eq.~\eqref{eq:ArzanoModel}. 

The second model we consider is the one by Breuer \textit{et al.}~\cite{Breuer:2008rh}, which takes the form 
\begin{equation}\label{eq:BreuerModel}
\partial_{t}\hat{\rho}(t)=-\frac{i}{\hbar}\left[\hat{H}_{0},\hat{\rho}(t)\right] - \tau_{B} \dfrac{1}{2\hbar^{2}} \left[\hat{K},\left[\hat{K},\hat{\rho}(t)\right]\right] \;.
\end{equation}
It describes fluctuations of the metric tensor around a Minkowski background, see Fig.~\ref{fig1}b, which is a way to model fluctuations of the light cone. The isotropic non-relativistic version of this proposal consists in adding to the background Euclidean metric a Kronecker delta multiplied by a parameter that is allowed to fluctuate around zero. This parameter and the time scale of its fluctuations are related to the quantum gravity scale. Since the norm of the spatial momentum is defined through the metric, the kinetic energy acquires an extra fluctuating contribution, which leads to a stochastic Schrödinger equation. The authors of \cite{Breuer:2008rh} then analyze the evolution of the corresponding averaged density matrix and find that the stochastic perturbation results in a Lindblad term as given in Eq.~\eqref{eq:BreuerModel}.

For a free particle, this model reproduces the energy-basis dephasing obtained by Blencowe~\cite{Blencowe:2012mp} for matter coupled to a thermal graviton bath at temperature $T$, with the identification $\tau_B = 2\hbar\,k_B T / E_P^2$ (see Appendix~\ref{suppsec:blencowe}), where $k_B$ is the Boltzmann constant and $E_P$ is the Planck energy. The correspondence is however limited to the free particle. Blencowe's dissipator is built from the full Hamiltonian, $[\hat H_0,[\hat H_0,\hat\rho]]$, and coincides with Eq.~\eqref{eq:BreuerModel} only when $\hat H_0=\hat K$. For a harmonic oscillator it would produce pure dephasing in the Fock basis and no heating, whereas the Breuer dissipator does not commute with $\hat H_0$ and induces the heating effect discussed below.

The third model we consider is the one by Petruzziello and Illuminati~\cite{Petruzziello:2020wkd}, which takes the form \cite{donadi_quantum_2025}
\begin{equation}\label{eq:IlluminatiModel}
\partial_{t}\hat{\rho}(t) = -\frac{i}{\hbar}[\hat{H}_0 + 4 a_P \hat{K}^2 \overline{\beta},\hat \rho] - \tau_I \dfrac{8 a_P^2}{\hbar^2} [\hat{K}^2,[\hat{K}^2, \hat \rho]] \;,
\end{equation}
where $a_P=m \ell_P^2/\hbar^2$, $\overline\beta$ the time average of the deformation parameter $\beta(t)$, and $\ell_P$ is the Planck length. 
The model is based on a generalized uncertainty principle (GUP), which is motivated by the possible existence of a minimal length set by the quantum gravity scale \cite{fadel_revisiting_2022}. 
Ref.~\cite{Petruzziello:2020wkd} makes this minimal length stochastic by allowing the GUP deformation parameter to undergo Gaussian white-noise fluctuations around a fixed, nonzero mean value, see Fig.~\ref{fig1}(c).
These random fluctuations of the GUP deformation parameter can be mapped onto an additional stochastic term in the effective Hamiltonian \cite{Petruzziello:2020wkd,donadi_quantum_2025}, which yields a Lindblad master equation \eqref{eq:IlluminatiModel} after averaging over the noise realizations.

In summary, the dissipation term of Eq.~\eqref{eq:ArzanoModel} measures the preservation of the relativity principle despite the existence of a quantum gravity scale, the one of Eq.~\eqref{eq:BreuerModel} describes the fluctuation of the spacetime metric due to quantum gravity effects, and the one of Eq.~\eqref{eq:IlluminatiModel} describes the possibility of the fluctuating minimal length in space due to quantum gravity.

\section{Free particle}

In this section we consider the case of a free particle. We will first review some results on the decoherence known from the literature \cite{Arzano:2022nlo,Breuer:2008rh,Petruzziello:2020wkd,donadi_quantum_2025}, to then extend them to effects that can be investigated through non-interferometric tests.

\subsection{Decoherence}

In the simple case of a free particle, the system Hamiltonian $\hat H_0$ is given by
\begin{equation}\label{eq:Hfree}
    \hat{H}_0=\hat{K}=\dfrac{\hat{p}^2}{2m} \;.
\end{equation}
As both the Hamiltonian and the dissipator are functions of $\hat p$, the state dynamics can be solved analytically in the momentum basis $\{\ket{p}\}$. By defining the matrix elements $\rho_{ab}(t):=\langle p_{a}|\hat{\rho}(t)|p_{b}\rangle$ and $\Delta E_k \equiv [(p^2_a/2m)^k-(p^2_b/2m)^k]$, we have for the Arzano model \cite{Arzano:2022nlo}
\begin{equation}\label{eq:ArzanoFreeDecay}
    \rho_{ab}(t) = \rho_{ab}(0) \; e^{-\frac{i}{\hbar} \Delta E_{1}t}e^{-(\tau_{A} c^2/2\hbar^{2}) (p_a - p_b)^{2}t} \;,
\end{equation}
for the Breuer model \cite{Breuer:2008rh,donadi_quantum_2025}
\begin{equation}\label{eq:BreuerFreeDecay}
    \rho_{ab}(t) = \rho_{ab}(0) \; e^{-\frac{i}{\hbar} \Delta E_{1}t}e^{-(\tau_{B}/2\hbar^{2}) (\Delta E_{1})^{2}t} \;,
\end{equation}
and for the Petruzziello--Illuminati model \cite{Petruzziello:2020wkd,donadi_quantum_2025}
\begin{equation}\label{eq:PetruzielloFreeDecay}
    \rho_{ab}(t) = \rho_{ab}(0) \; e^{-\frac{i}{\hbar}\left(\Delta E_{1}+4a_{P}\Delta E_{2}\overline{\beta}\right)t}
\; e^{-(8 a_{P}^{2}\tau_I /\hbar^{2})\left( \Delta E_2 \right)^{2} t} \;.
\end{equation}
The real exponential allows us to identify the associated decoherence rates. Defining $\Delta p=(p_a-p_b)$ and $\overline{p}=(p_a+p_b)/2$, we have
\begin{subequations}
\begin{align}
    & \Gamma_A =  \dfrac{\tau_{A} c^2}{2\hbar^{2}} \Delta p^{2} \;,\qquad \Gamma_B = \dfrac{\tau_{B}}{2\hbar^{2}} \dfrac{\Delta p^2 \overline{p}^2}{m^2} \;, \\
    &\qquad \Gamma_I = \dfrac{2 a_{P}^{2}\tau_I }{\hbar^{2}} \dfrac{\Delta p^2 \overline{p}^2 (p_a^2+p_b^2)^2}{m^4} \;,
\end{align}
\end{subequations}
respectively.
Interestingly, note that for $p_a=-p_b$ only $\Gamma_A\neq 0$.
If we consider $p_{a,b}=\overline{p}\pm\Delta p/2$ and expand for small $\Delta p$, we obtain the scalings
$\Gamma_A\sim \Delta p^2$, $\Gamma_B\sim \overline{p}^2 \Delta p^2$ and $\Gamma_I \sim \overline{p}^6 \Delta p^2$, all quadratic in $\Delta p$ but with different $\overline{p}$ dependences.

\subsection{Position diffusion}

In the absence of a potential, both the Hamiltonian and the dissipator are functions of $\hat p^n$ only.
This implies that energy is conserved, $d \avg{\hat H_0}_t/dt = 0$, but also that there is no change in momentum.  
In general, given a Hermitian jump operator $\hat C$ the equation for the evolution of the expectation value of an operator $\hat O$ is
\begin{equation}\label{eq:eomOpO}
 \dfrac{d}{dt} \avg{\hat O}_t = \dfrac{i}{\hbar}\avg{[\hat H_0,\hat O]}_t - \dfrac{1}{2} \avg{[\hat C,[\hat C,\hat O]]}_t  \;.
\end{equation}
If $\hat C = f(\hat p)$ is a function only of $\hat p$, as in the three models we are considering, then $[\hat C,[\hat C,\hat x]]=0$, $[\hat C,[\hat C,g(\hat p)]]=0$ and $[\hat C,[\hat C,(\hat x\hat p+\hat p\hat x)]]=0$. The equations of motion for the first moments of momentum, position, and for their correlations, are then found to coincide with those of a free particle
\begin{equation}\label{eq:freeXPeom}
\dfrac{d}{dt} \avg{\hat p}_t = 0 \;,\quad \dfrac{d}{dt} \avg{\hat x}_t = \dfrac{1}{m}\avg{\hat p}_t \;,\quad \dfrac{d}{dt} \avg{\hat x\hat p + \hat p\hat x}_t = \dfrac{2}{m} \avg{\hat p^2}_t 
\end{equation}
For the Petruzziello--Illuminati model, Eqs.~(\ref{eq:eomOpO},\ref{eq:freeXPeom}) hold once the coherent correction $4a_P\overline{\beta}\hat K^2$ in Eq.~\eqref{eq:IlluminatiModel} is neglected. Retaining it adds $(4a_P\overline{\beta}/m^2)\avg{\hat p^3}_t$ to $d\avg{\hat x}_t/dt$ and $(8a_P\overline{\beta}/m^2)\avg{\hat p^4}_t$ to $d\avg{\hat x\hat p+\hat p\hat x}_t/dt$. Since these moments are conserved, the correction would only add linear drifts, suppressed relative to the free terms by $\overline{\beta}\,\ell_P^2\avg{\hat p^2}/\hbar^2$, and it does not lead to decoherence.
Integrating the equations of motion Eqs.~\eqref{eq:freeXPeom} gives
\begin{subequations}\label{eq:freeXPsol}
\begin{align}
    \avg{\hat p}_t &= \avg{\hat p}_0 \\
    \avg{\hat x}_t &= \avg{\hat x}_0 + \dfrac{t}{m} \avg{\hat p}_0 \\
    \avg{\hat x\hat p + \hat p\hat x}_t &= \avg{\hat x\hat p + \hat p\hat x}_0 + \dfrac{2 t}{m} \avg{\hat p^2}_0
\end{align}
\end{subequations}
The second moment $\langle\hat x^2\rangle$, however, is affected, since $[\hat C,[\hat C,\hat x^2]]\neq0$. So a measurement of $\langle \hat x^2\rangle$ will vary in time.
Using Eq.~\eqref{eq:eomOpO} with $\hat H_0=\hat p^2/2m$ and $\hat C=f(\hat p)$, we find the equation of motion 
\begin{equation}
\frac{d}{dt}\langle \hat x^2\rangle_t
=\frac{1}{m}\langle \hat x\hat p+\hat p\hat x\rangle_t
+\hbar^2\left\langle\big(f'(\hat p)\big)^2\right\rangle_t,
\end{equation}
so the dissipator induces an additional, nonnegative growth of $\langle \hat x^2\rangle_t$, a position diffusion, even though the momentum distribution, and hence energy, is unchanged.

\subsection{General momentum dissipator}

Consider again the free particle case $\hat{H}_{0}=\hat p^2/2m$, now with a collapse operator proportional to $\hat p^{j}$, $j\in\{1,2,\dots\}$,
\begin{equation}
\partial_{t}\hat{\rho}(t)=-\frac{i}{\hbar}\left[\hat{H}_{0},\hat{\rho}(t)\right] - \dfrac{\tau}{\mathcal{N}} \left[\hat{p}^{j},\left[\hat{p}^{j},\hat{\rho}(t)\right]\right] \;.
\end{equation}
The normalization constant must have dimensions $\mathcal{N}\sim(\text{momentum}^{2j})(\text{time}^2)$. 
The three models above correspond to $j=1$ (Arzano), $j=2$ (Breuer) and $j=4$ (Petruzziello--Illuminati). See also the case of stochastic modified dispersion relations discussed in Ref.~\cite{Lobo:2025mze}.

The equations of motion for $\avg{\hat x}$, $\avg{\hat p}$ and $\avg{\hat x\hat p+\hat p\hat x}$ are given by Eq.~\eqref{eq:freeXPeom}, with solutions in Eq.~\eqref{eq:freeXPsol}. Using $[\hat p^{j},[\hat p^{j},\hat x^2]]=-2\hbar^2 j^2\hat p^{2j-2}$ we have
\begin{equation}
\dfrac{d}{dt} \avg{\hat x^2}_t = \dfrac{1}{m}\avg{\hat x\hat p + \hat p\hat x}_t + \dfrac{2 \tau \hbar^2 j^2}{\mathcal{N}}\avg{\hat p^{2j-2}}_t \;,
\end{equation}
with solution
\begin{equation}\label{eq:heating_free}
\avg{\hat x^2}_t = \avg{\hat x^2}_0 + \dfrac{t}{m} \avg{\hat x\hat p + \hat p\hat x}_0 + \dfrac{t^2}{m^2} \avg{\hat p^2}_0 + \dfrac{2 \tau \hbar^2 j^2}{\mathcal{N}}\avg{\hat p^{2j-2}}_0\, t \;.
\end{equation}
This means that observing a linear diffusion for short time evolutions can set a bound on the product $\tau j^2$. Moreover, the specific $j$ dependence could be identified by preparing initial states with different momentum distributions.

This result has a transparent interpretation in terms of a stochastic unraveling. The master equation above is the ensemble average of the stochastic Schr\"odinger equation
\begin{equation}
i\hbar\frac{d}{dt}|\psi_t\rangle=\left[\frac{\hat p^2}{2m}+\hbar\sqrt{\frac{2\tau}{\mathcal N}}\,\hat p^{j}\,\xi(t)\right]|\psi_t\rangle,
\end{equation}
with $\xi(t)$ the real Gaussian white noise satisfying $\mathbb E[\xi(t)]=0$ and $\mathbb E[\xi(t)\xi(t')]=\delta(t-t')$. 
The corresponding Heisenberg equations,
\begin{equation}
\dot{\hat x}=\frac{\hat p}{m}+j\hbar\sqrt{\frac{2\tau}{\mathcal N}}\,\hat p^{j-1}\,\xi(t),\qquad \dot{\hat p}=0,
\end{equation}
show that the noise adds a random, momentum-dependent term to the (group) velocity $\dot{\hat x}=\partial \hat H/\partial \hat p$, with $\hat H$ the stochastic Hamiltonian in brackets above.
Since $\hat p$ is conserved, the momentum distribution and the energy are unchanged, while each momentum component performs a random walk in position, $\hat x_t=\hat x_0+\hat pt/m+j\hbar\sqrt{2\tau/\mathcal N}\,\hat p^{j-1}W_t$, with $W_t$ a Wiener process. Averaging $\hat x_t^2$ reproduces Eq.~\eqref{eq:heating_free}, whose last term is thus a position diffusion with a momentum-dependent coefficient. The same noise accounts for the decoherence in the momentum basis, since two components $p_a$ and $p_b$ acquire the relative random phase $\sqrt{2\tau/\mathcal N}\,(p_a^j-p_b^j)W_t$, whose average gives a decay at the rate $(\tau/\mathcal N)(p_a^j-p_b^j)^2$. For $j=1$ the noise is additive and the diffusion is independent of the state, whereas for $j\geq2$ it is multiplicative and grows with the momentum.

\section{Harmonic oscillator}

In this section we want to generalize the previous free particle results to the case of a harmonic oscillator. The latter is particularly relevant as it describes experimental platforms that can find application in tests of decoherence mechanisms.
For simplicity, we consider a harmonic oscillator in one dimension and leave higher-dimension extensions as straightforward generalizations.
The system Hamiltonian $\hat H_0$ is now
\begin{subequations}\label{eq:HHOdef}
\begin{align}
    \hat H_0 &= \hat K + \hat U = \dfrac{\hat p^2}{2m} + \dfrac{1}{2} m \omega^2 \hat x^2   \label{eq:H0def}\\
    &= \hbar\omega\left( \hat n + \dfrac{1}{2} \right) \;,
\end{align}
\end{subequations}
where $\omega$ is the oscillator frequency, $\hat x$ the position operator, and $\hat n$ the number operator.
The total oscillator energy is
\begin{equation}\label{eq:defEt}
E_t \equiv \langle \hat H_0\rangle_t
= \langle \hat K\rangle_t+\langle \hat U\rangle_t
= \hbar\omega\left(\langle \hat n\rangle_t+\frac{1}{2}\right),
\end{equation}
where the factor \(1/2\) is the zero-point energy contribution.

As experimental realizations of harmonic oscillators always suffer from energy relaxation (amplitude damping) at some rate $\gamma$, we include in the master equations~(\ref{eq:ArzanoModel},\ref{eq:BreuerModel},\ref{eq:IlluminatiModel}) the term
\begin{equation}\label{eq:dissZeroT}
\gamma\mathcal D[\hat a]\hat \rho ,\qquad \mathcal D[\hat a]\hat \rho =\left( \hat a \hat \rho \hat a^\dagger - \frac12\{\hat a^\dagger \hat a, \hat\rho\} \right)\;,
\end{equation}
with $\hat a$ and $\hat a^\dagger$ are the usual (bosonic) ladder operators of the harmonic oscillator.
These are defined as $\hat a=\sqrt{m\omega/2\hbar}\,(\hat x+i\hat p/m\omega)$ and $\hat a^\dagger=\sqrt{m\omega/2\hbar}\,(\hat x-i\hat p/m\omega)$, so that $[\hat a,\hat a^\dagger]=1$ and $\hat n=\hat a^\dagger\hat a$, consistently with Eq.~\eqref{eq:H0def}.

While Eq.~\eqref{eq:dissZeroT} describes amplitude damping of the oscillator through its coupling to a zero-temperature bath, when discussing more realistic experimental scenarios we will occasionally consider a finite-temperature bath. This is described by the two dissipators $\gamma(n_{\rm th}+1)\mathcal D[\hat a]\hat\rho+\gamma n_{\rm th}\mathcal D[\hat a^\dagger]\hat\rho$, where $n_{\rm th}$ is the thermal occupation of the bath.
They add a term $\gamma n_{\rm th}$ to the equation for $\partial_t\langle\hat n\rangle_t$. For the Arzano model, this term simply adds to the gravity-induced one, whereas for the Breuer model the thermal phonons are amplified by the kinetic-energy dependence.
We will not consider here other forms of coupling to the environment, such as those resulting in pure dephasing through a term $\gamma_\phi\mathcal D[\hat a^\dagger \hat a]\hat\rho$ in the master equation.

\subsection{Decoherence}
\label{sec:HOdecoherence}

We first investigate how the three gravity-related decoherence models affect different states of the harmonic oscillator. In particular, we consider the initial states
\begin{equation}
\hat{\rho}_0(0)=|0\rangle\langle0|,\quad \hat{\rho}_1(0)=|1\rangle\langle1|,\quad \hat{\rho}_{0+1}(0)=|\psi_{0+1}\rangle\langle\psi_{0+1}|,
\end{equation}
where $|\psi_{0+1}\rangle=(|0\rangle+|1\rangle)/\sqrt{2}$.
The first two states allow us to study the evolution of Fock-state populations, while the third directly probes the coherence between the states $|0\rangle$ and $|1\rangle$. In the absence of the gravity-related terms, energy relaxation gives $\langle0|\hat{\rho}_0(t)|0\rangle=1$, $\langle1|\hat{\rho}_1(t)|1\rangle=e^{-\gamma t}$, and $\langle0|\hat{\rho}_{0+1}(t)|1\rangle=e^{i\omega t}e^{-\gamma t/2}/2$.

\textit{Arzano model.--} For the Arzano model, the harmonic-oscillator dynamics with energy relaxation can be solved exactly, as shown in Appendix~\ref{Arzano_exact}. It is convenient to introduce the characteristic oscillator rate
\begin{equation}
\Gamma_A^{\rm HO}\equiv\frac{m\omega c^2\tau_A}{\hbar}.
\label{eq:GammaAHO}
\end{equation}
To leading order in $\Gamma_A^{\rm HO}$ and in the regime $\gamma t\ll1\ll\omega t$, where non-secular terms of relative order $\Gamma_A^{\rm HO}/\omega$ can be neglected, the coherence of the state $\hat{\rho}_{0+1}$ evolves as
\begin{equation}
\langle0|\hat{\rho}_{0+1}(t)|1\rangle\simeq\frac{1}{2}e^{i\omega t}e^{-\gamma t/2}\left(1-\Gamma_A^{\rm HO}t\right),
\label{eq:ArzanoHOcoherence}
\end{equation}
while the Fock-state probabilities are
\begin{equation}
\langle0|\hat{\rho}_0(t)|0\rangle\simeq1-\frac{1}{2}\Gamma_A^{\rm HO}t,\qquad \langle1|\hat{\rho}_1(t)|1\rangle\simeq e^{-\gamma t}\left(1-\frac{3}{2}\Gamma_A^{\rm HO}t\right).
\label{eq:ArzanoHOpopulations}
\end{equation}
Since $\hat p$ has no diagonal elements in the Fock basis, the coherence-loss rate is simply the mean of the two population-loss rates. The exact solution in Appendix~\ref{Arzano_exact} shows that these expressions describe only the initial deviation from standard dynamics. At longer times the Arzano dissipator continuously redistributes population among Fock states, so that its interplay with energy relaxation cannot in general be just absorbed into a constant correction to $\gamma$. 


\textit{Breuer model.--} For this model, the corresponding harmonic-oscillator dynamics was derived in Ref.~\cite{donadi_quantum_2025}. Defining
\begin{equation}
\Gamma_B^{\rm HO}\equiv\tau_B\omega^2,
\label{eq:GammaBHO}
\end{equation}
to leading order in $\Gamma_B^{\rm HO}$ and in the same regime $\gamma t\ll1\ll\omega t$ as for the Arzano model, the $|0\rangle$--$|1\rangle$ coherence evolves as
\begin{equation}
\langle0|\hat{\rho}_{0+1}(t)|1\rangle\simeq\frac{1}{2}e^{i\omega t}e^{-\gamma t/2}\left(1-\frac{3}{8}\Gamma_B^{\rm HO}t\right),
\label{eq:BreuerHOcoherence}
\end{equation}
while the Fock-state survival probabilities are
\begin{equation}
\langle0|\hat{\rho}_0(t)|0\rangle\simeq1-\frac{1}{8}\Gamma_B^{\rm HO}t,\quad \langle1|\hat{\rho}_1(t)|1\rangle\simeq e^{-\gamma t}\left(1-\frac{3}{8}\Gamma_B^{\rm HO}t\right).
\label{eq:BreuerHOpopulations}
\end{equation}
These expressions are obtained perturbatively in the metric-fluctuation term \cite{donadi_quantum_2025}. As in the Arzano case, the decay of $\langle0|\hat{\rho}_0(t)|0\rangle$ is particularly noteworthy: ordinary zero-temperature amplitude damping leaves the oscillator ground state invariant, so any reduction of its population is entirely induced by the additional decoherence mechanism. In this sense, ground-state depletion can also be viewed as a heating signature of the model, although the fact that here phonons are created in pairs implies that the stationary state is not thermal.

\textit{Petruzziello--Illuminati model.--} For the stochastic-GUP model, the corresponding harmonic-oscillator calculation was performed in Ref.~\cite{donadi_quantum_2025}, using a rotating-wave approximation for the $\overline{\beta}$-dependent Hamiltonian correction and treating the decoherence term perturbatively. In terms of the characteristic oscillator rate
\begin{equation}
\Gamma_I^{\rm HO}\equiv8a_P^2\tau_I\hbar^2\omega^4,
\label{eq:GammaIHO}
\end{equation}
with $a_P$ is defined below \eqref{eq:IlluminatiModel}, and in the same regime $\gamma t\ll1\ll\omega t$ as for the two previous models, the coherence of $\hat{\rho}_{0+1}$ evolves as
\begin{equation}
\langle0|\hat{\rho}_{0+1}(t)|1\rangle\simeq\frac{1}{2}e^{-\frac{i}{\hbar}(E_0^I-E_1^I)t}e^{-\gamma t/2}\left(1-\frac{30}{8}\Gamma_I^{\rm HO}t\right),
\label{eq:IlluminatiHOcoherence}
\end{equation}
where $E_n^I=\hbar\omega(n+\tfrac12)+\tfrac32a_P\overline{\beta}(\hbar\omega)^2(n^2+n+\tfrac12)$ is the oscillator energy including the coherent $\overline{\beta}$-dependent correction \footnote{The coefficient $\tfrac32$ follows from $4a_P\overline\beta\langle n|\hat K^2|n\rangle$. The value $\tfrac38$ in Ref.~\cite{donadi_quantum_2025} omits the prefactor $4$ of the coherent correction, which affects only the phase and not the decoherence rates.}. The latter modifies the phase evolution but is distinct from the decoherence induced by fluctuations of the GUP parameter. For the initial states $|0\rangle$ and $|1\rangle$, one finds
\begin{equation}
\langle0|\hat{\rho}_0(t)|0\rangle\simeq1-\frac{6}{8}\Gamma_I^{\rm HO}t,\qquad \langle1|\hat{\rho}_1(t)|1\rangle\simeq e^{-\gamma t}\left(1-\frac{45}{8}\Gamma_I^{\rm HO}t\right).
\label{eq:IlluminatiHOpopulations}
\end{equation}

A common feature of all three models is that, contrary to the free-particle case, harmonic-oscillator energy eigenstates are not eigenstates of the corresponding collapse operators, since $[\hat H_0, \hat p^k]\neq 0$. Consequently, the gravity-related terms affect not only coherences between different Fock states, but also the populations of individual Fock states. In particular, all three models predict a departure of an initially prepared ground state from $|0\rangle$, providing a non-interferometric signature that is absent for zero-temperature amplitude damping.

The different momentum dependence of the three collapse operators is reflected in markedly different characteristic oscillator rates Eqs.~(\ref{eq:GammaAHO},\ref{eq:GammaBHO},\ref{eq:GammaIHO}), which we summarize here:
\begin{equation}\label{eq:HOdecoherenceRates}
\Gamma_A^{\rm HO}=\frac{m\omega c^2\tau_A}{\hbar},\qquad \Gamma_B^{\rm HO}=\tau_B\omega^2,\qquad \Gamma_I^{\rm HO}=8a_P^2\tau_I\hbar^2\omega^4.
\end{equation}
Since $a_P\propto m$, these scale respectively as $m\omega$, $\omega^2$, and $m^2\omega^4$ for a single particle and fixed model parameters. 
For composite oscillators, the mass dependence is modified by the correlations of the fluctuations and by the relative motion of the atoms, as discussed later.
Measurements of both Fock-state populations and coherences, performed for oscillators with different masses or frequencies, therefore provide complementary observables with which the three phenomenological models can in principle be constrained and distinguished.

The decoherence corrections presented in this subsection are evaluated to leading order in the gravity-induced noise strength. For the Arzano model, exact populations and coherences are derived in Appendix~\ref{Arzano_exact}, whereas for the Breuer model only the second moments admit an exact solution, given in Appendix~\ref{app:BreuerExact}. For the Petruzziello--Illuminati model, the moment equations do not close, and no exact solution is available.
These perturbative expressions describe the initial deviations from standard quantum dynamics but do not, in general, capture the full long-time behavior. In the next subsection, we derive the exact energy dynamics for the Arzano and Breuer models, revealing qualitatively different heating mechanisms, namely linear energy growth in the absence of damping for the former, and exponential growth for the latter when mechanical damping is sufficiently weak. For the Petruzziello--Illuminati model, the heating rate depends on $\langle\hat K^3\rangle$ and thus grows strongly with the excitation of the oscillator.

\subsection{Heating}
\label{sec:heating}

We now investigate the energy evolution induced by the gravity-induced modifications, also in the presence of amplitude damping. For a one-dimensional harmonic oscillator, the expectation value of the total energy is Eq.~\eqref{eq:defEt}, where, for compactness, we introduce
\begin{equation}
K_t\equiv\langle\hat K\rangle_t,\qquad U_t\equiv\langle\hat U\rangle_t,\qquad S_t\equiv\langle\hat S\rangle_t,
\end{equation}
with $\hat S=\hat x\hat p+\hat p\hat x$.

The Arzano and Breuer models lead to qualitatively different heating mechanisms. In Arzano's case the deformation acts as a constant source of energy, leading to a linear increase in the absence of dissipation. In Breuer's case, instead, the deformation-induced contribution is proportional to the kinetic energy itself and may generate an exponential instability. We analyze these two behaviors separately, before turning to the Petruzziello--Illuminati model, whose moment equations do not close.

\subsubsection{Arzano model}
\label{sec:ArzanoHeating}

For the Arzano model including amplitude damping, the equations of motion for the second moments are
\begin{subequations}
\label{eq:ArzanoMoments}
\begin{align}
\dot K_t&=-\frac{\omega^2}{2}S_t+\gamma\left(\frac{\hbar\omega}{4}-K_t\right),\\
\dot U_t&=\frac{\omega^2}{2}S_t+\frac{\hbar\omega}{2}\Gamma_A^{\rm HO}+\gamma\left(\frac{\hbar\omega}{4}-U_t\right),\\
\dot S_t&=4K_t-4U_t-\gamma S_t.
\end{align}
\end{subequations}
The Arzano term therefore acts as a constant source in the potential-energy equation, while leaving the equation for the kinetic energy unchanged.

Adding the first two equations in Eq.~\eqref{eq:ArzanoMoments}, the mixed-moment contribution cancels and the total energy obeys the closed equation
\begin{equation}
\dot E_t=\frac{\hbar\omega}{2}\Gamma_A^{\rm HO}+\gamma\left(\frac{\hbar\omega}{2}-E_t\right).
\label{eq:ArzanoHeatingEnergy}
\end{equation}
Thus, in the absence of damping, the Arzano dissipator injects energy into the oscillator at the constant rate $\hbar\omega\Gamma_A^{\rm HO}/2$.

Using Eq.~\eqref{eq:defEt}, Eq.~\eqref{eq:ArzanoHeatingEnergy} becomes
\begin{equation}
\partial_t\langle\hat n\rangle_t=-\gamma\langle\hat n\rangle_t+\frac{\Gamma_A^{\rm HO}}{2},
\label{eq:ArzanoHeatingNumber}
\end{equation}
whose solution for an arbitrary initial occupation is
\begin{equation}
\langle\hat n\rangle_t=\langle\hat n\rangle_0e^{-\gamma t}+\frac{\Gamma_A^{\rm HO}}{2\gamma}\left(1-e^{-\gamma t}\right).
\label{eq:ArzanoNumberSolution}
\end{equation}
In the absence of amplitude damping, the energy therefore increases linearly at all times,
\begin{equation}
E_t=E_0+\frac{\hbar\omega}{2}\Gamma_A^{\rm HO}t,
\label{eq:ArzanoLinearHeating}
\end{equation}
in contrast with the free-particle case, where the same dissipator produces position diffusion but leaves the kinetic energy unchanged. With damping, the same linear growth holds at short times, $\gamma t\ll1$, and for an oscillator initially in the ground state
\begin{equation}
\langle\hat n\rangle_t\simeq\frac{\Gamma_A^{\rm HO}}{2}t.
\label{eq:ArzanoHeatingShort}
\end{equation}
This is consistent with the ground-state depletion of Eq.~\eqref{eq:ArzanoHOpopulations}, since to first order the loss of ground-state population corresponds to the creation of a single phonon. At long times, the energy injection is balanced by amplitude damping, and the oscillator approaches
\begin{equation}
\langle\hat n\rangle_\infty=\frac{\Gamma_A^{\rm HO}}{2\gamma},\qquad E_\infty=\frac{\hbar\omega}{2}\left(1+\frac{\Gamma_A^{\rm HO}}{\gamma}\right).
\label{eq:ArzanoHeatingStationary}
\end{equation}
The Arzano parameter can therefore be constrained either from the anomalous short-time heating rate or from the excess stationary occupation.

Although the total energy follows the simple dynamics of Eq.~\eqref{eq:ArzanoHeatingEnergy}, the kinetic and potential contributions also exhibit an oscillatory exchange. Combining Eqs.~\eqref{eq:ArzanoMoments} gives
\begin{equation}
\ddot S_t+2\gamma\dot S_t+(\gamma^2+4\omega^2)S_t=-2\hbar\omega\Gamma_A^{\rm HO}.
\label{eq:ArzanoMixedMoment}
\end{equation}
The characteristic roots of Eq.~\eqref{eq:ArzanoMixedMoment} are $-\gamma\pm2i\omega$, so that
\begin{align}
S_t&=S_{\rm ss}+e^{-\gamma t}\left[A\cos(2\omega t)+B\sin(2\omega t)\right],\\
S_{\rm ss}&=-\frac{2\hbar\omega\Gamma_A^{\rm HO}}{\gamma^2+4\omega^2},
\end{align}
with $A$ and $B$ fixed by the initial conditions. From the last of Eqs.~\eqref{eq:ArzanoMoments}, $K_t-U_t=(\dot S_t+\gamma S_t)/4$ contains the same oscillating terms, whereas $K_t+U_t=E_t$ obeys the closed Eq.~\eqref{eq:ArzanoHeatingEnergy} and does not oscillate. The kinetic and potential energies therefore oscillate in antiphase at $2\omega$ with an envelope $e^{-\gamma t}$, describing the exchange of energy between the two sectors, and these oscillations cancel in their sum.

\subsubsection{Breuer model}
\label{sec:BreuerHeating}
For the Breuer model, the equations of motion for the second moments are
\begin{subequations}
\label{eq:BreuerMoments}
\begin{align}
\dot K_t&=-\frac{\omega^2}{2}S_t+\gamma\left(\frac{\hbar\omega}{4}-K_t\right),\\
\dot U_t&=\frac{\omega^2}{2}S_t+\Gamma_B^{\rm HO}K_t+\gamma\left(\frac{\hbar\omega}{4}-U_t\right),\\
\dot S_t&=4K_t-4U_t-\gamma S_t.
\end{align}
\end{subequations}
As for the Arzano model, the gravity-related term enters only the equation for the potential energy, but it is now proportional to the instantaneous kinetic energy rather than constant.

Adding the first two equations in Eq.~\eqref{eq:BreuerMoments}, the energy is
\begin{equation}
\dot E_t=\Gamma_B^{\rm HO}K_t+\gamma\left(\frac{\hbar\omega}{2}-E_t\right),
\label{eq:BreuerHeatingEnergy}
\end{equation}
or equivalently, using Eq.~\eqref{eq:defEt},
\begin{equation}
\partial_t\langle\hat n\rangle_t=-\gamma\langle\hat n\rangle_t+\Gamma_B^{\rm HO}\frac{K_t}{\hbar\omega}.
\label{eq:BreuerHeatingNumber}
\end{equation}
The Breuer contribution therefore does not define a state-independent heating rate, since the energy injection grows with the kinetic energy of the oscillator, as in a parametric process. For an oscillator initially in the ground state, $K_0=\hbar\omega/4$, and the initial heating rate is
\begin{equation}
\left.\partial_t\langle\hat n\rangle_t\right|_{t=0}=\frac{\Gamma_B^{\rm HO}}{4}.
\label{eq:BreuerInitialHeating}
\end{equation}
This is consistent with the ground-state depletion of Eq.~\eqref{eq:BreuerHOpopulations}, since the collapse operator $\hat K\propto\hat p^2$ couples $|0\rangle$ to $|2\rangle$, so that transitions occurring at rate $\Gamma_B^{\rm HO}/8$ create phonons in pairs.

At longer times, the dynamics differs qualitatively from the linear heating of the Arzano model. Its exact solution is derived in Appendix~\ref{app:BreuerExact}, and here we discuss its stability and stationary state. The homogeneous part of Eq.~\eqref{eq:BreuerMoments} reads
\begin{equation}
\frac{d}{dt}
\begin{pmatrix}
K_t\\ U_t\\ S_t
\end{pmatrix}
=
\left[
\begin{pmatrix}
0&0&-\omega^2/2\\
\Gamma_B^{\rm HO}&0&\omega^2/2\\
4&-4&0
\end{pmatrix}
-\gamma \id
\right]
\begin{pmatrix}
K_t\\ U_t\\ S_t
\end{pmatrix},
\label{eq:BreuerMatrix}
\end{equation}
where $\id$ is the identity matrix. The eigenvalues of the undamped matrix are the roots of
\begin{equation}
r^3+4\omega^2r-2\Gamma_B^{\rm HO}\omega^2=0.
\label{eq:rBreuer}
\end{equation}
Since the left-hand side increases monotonically with $r$ and is negative at $r=0$, there is the unique real root
\begin{equation}
r=\frac{4\omega}{\sqrt3}\,\sinh\left[\frac13\,\mathrm{arsinh}\left(\frac{3\sqrt3}{8}\,\frac{\Gamma_B^{\rm HO}}{\omega}\right)\right],
\label{eq:rBreuerExplicit}
\end{equation}
which is positive and reduces to $r\simeq\Gamma_B^{\rm HO}/2$ for $\Gamma_B^{\rm HO}\ll\omega$. The other two roots are $-r/2\pm i\Omega$, with
\begin{equation}
\Omega=\frac{1}{2}\sqrt{3r^2+16\omega^2}.
\label{eq:OmegaBreuer}
\end{equation}
Amplitude damping shifts all three eigenvalues by $-\gamma$, so that the characteristic exponents become $r-\gamma$ and $-\gamma-r/2\pm i\Omega$. The oscillatory pair always decays, whereas the nonoscillatory mode decays only for $\gamma>r$. 
The competition between the Breuer-induced heating and mechanical damping therefore defines a stability threshold at the critical damping rate $\gamma_{\rm c} \equiv r$. By the monotonicity of Eq.~\eqref{eq:rBreuer}, the stability condition $\gamma>\gamma_{\rm c}$ is equivalent to
\begin{equation}
2\Gamma_B^{\rm HO}\omega^2<\gamma\left(\gamma^2+4\omega^2\right),
\label{eq:BreuerStabilityCondition}
\end{equation}
which is the condition for the existence of a stationary state with finite second moments.

In the stable regime, setting the time derivatives in Eq.~\eqref{eq:BreuerMoments} to zero gives
\begin{subequations}
\label{eq:BreuerStationary}
\begin{align}
K_\infty&=\frac{\gamma\hbar\omega(\gamma^2+4\omega^2)}{4[\gamma(\gamma^2+4\omega^2)-2\Gamma_B^{\rm HO}\omega^2]},\\
U_\infty&=\frac{\gamma\hbar\omega(\gamma^2+\gamma\Gamma_B^{\rm HO}+4\omega^2)}{4[\gamma(\gamma^2+4\omega^2)-2\Gamma_B^{\rm HO}\omega^2]},\\
S_\infty&=-\frac{\gamma\hbar\omega\Gamma_B^{\rm HO}}{\gamma(\gamma^2+4\omega^2)-2\Gamma_B^{\rm HO}\omega^2},
\end{align}
\end{subequations}
which reduce to $K_\infty=U_\infty=\hbar\omega/4$ and $S_\infty=0$ for $\Gamma_B^{\rm HO}\rightarrow0$, as expected for zero-temperature amplitude damping. The corresponding excess stationary occupation is
\begin{equation}
\langle\hat n\rangle_\infty=\frac{\Gamma_B^{\rm HO}(\gamma^2+4\omega^2)}{4[\gamma(\gamma^2+4\omega^2)-2\Gamma_B^{\rm HO}\omega^2]}.
\label{eq:BreuerStationaryNumber}
\end{equation}
The common denominator factorizes as $\gamma(\gamma^2+4\omega^2)-2\Gamma_B^{\rm HO}\omega^2=(\gamma-\gamma_{\rm c})(\gamma^2+\gamma\gamma_{\rm c}+\gamma_{\rm c}^2+4\omega^2)$, which is positive precisely in the stable regime, so that the stationary second moments diverge as $(\gamma-\gamma_{\rm c})^{-1}$ when the threshold is approached. Far from it, for $\gamma\gg\gamma_{\rm c}$, the stationary occupation reduces to
\begin{equation}
\langle\hat n\rangle_\infty\simeq\frac{\Gamma_B^{\rm HO}}{4\gamma},
\label{eq:BreuerStationaryWeak}
\end{equation}
which corresponds to amplitude damping balancing the initial heating rate of Eq.~\eqref{eq:BreuerInitialHeating}.

The origin of the instability becomes transparent in the experimentally relevant regime $\gamma,\Gamma_B^{\rm HO}\ll\omega$, where $r\simeq\Gamma_B^{\rm HO}/2$, $\Omega\simeq2\omega$, and the stability condition reduces to $\gamma>\Gamma_B^{\rm HO}/2$. Averaging over the rapid exchange between kinetic and potential energy, for which $K_t\simeq E_t/2$, Eq.~\eqref{eq:BreuerHeatingNumber} becomes
\begin{equation}
\partial_t\langle\hat n\rangle_t\simeq-\left(\gamma-\frac{\Gamma_B^{\rm HO}}{2}\right)\langle\hat n\rangle_t+\frac{\Gamma_B^{\rm HO}}{4}.
\label{eq:BreuerSecularNumber}
\end{equation}
For $\Gamma_B^{\rm HO}>2\gamma$, the effective damping changes sign and the occupation grows exponentially. In particular, in the absence of amplitude damping, the energy grows asymptotically as
\begin{equation}
E_t\propto \exp\left( \dfrac{\Gamma_B^{\rm HO}t}{2} \right).
\label{eq:BreuerExponentialHeating}
\end{equation}

The physical distinction from the Arzano model is therefore clear. The Arzano dissipator acts as an additive, state-independent energy source and produces linear heating in the absence of damping. The Breuer dissipator instead injects energy at a rate proportional to the kinetic energy, a positive feedback that leads to exponential growth when the damping rate falls below the threshold $\gamma_{\rm c}$.

\subsubsection{Petruzziello--Illuminati model}
\label{sec:IlluminatiHeating}

The Petruzziello--Illuminati model differs qualitatively from the previous two because its collapse operator is proportional to $\hat K^2$. Starting from Eq.~\eqref{eq:IlluminatiModel} and including amplitude damping, the second moments obey
\begin{subequations}
\label{eq:IlluminatiMoments}
\begin{align}
\dot K_t&=-\frac{\omega^2}{2}S_t+\gamma\left(\frac{\hbar\omega}{4}-K_t\right),\\
\dot U_t&=\frac{\omega^2}{2}S_t+2a_P\overline{\beta}\omega^2\langle\{\hat K,\hat S\}\rangle_t\notag\\
&+64a_P^2\tau_I\omega^2\langle\hat K^3\rangle_t+\gamma\left(\frac{\hbar\omega}{4}-U_t\right),\\
\dot S_t&=4K_t-4U_t+32a_P\overline{\beta}\langle\hat K^2\rangle_t-\gamma S_t.
\end{align}
\end{subequations}
Unlike for the Arzano and Breuer models, higher moments enter in these equations, which therefore do not form a closed system. Adding the first two of them, the bare oscillator energy satisfies
\begin{equation}
\dot E_t=2a_P\overline{\beta}\omega^2\langle\{\hat K,\hat S\}\rangle_t+64a_P^2\tau_I\omega^2\langle\hat K^3\rangle_t+\gamma\left(\frac{\hbar\omega}{2}-E_t\right).
\label{eq:IlluminatiEnergyFull}
\end{equation}
The first term originates from the deterministic GUP correction proportional to $\overline{\beta}$ and represents a coherent modification of the dynamics. The second term originates from the fluctuations of the deformation parameter and is responsible for the anomalous heating.

For consistency with Sec.~\ref{sec:HOdecoherence} and Ref.~\cite{donadi_quantum_2025}, we apply the rotating-wave approximation to the coherent GUP correction while retaining the full $\hat K^2$ dependence of the decoherence term. The resulting coherent Hamiltonian is a function of $\hat n$ and commutes with $\hat H_0$, so that it does not change the mean occupation, and Eq.~\eqref{eq:IlluminatiEnergyFull} reduces to
\begin{equation}
\dot E_t=64a_P^2\tau_I\omega^2\langle\hat K^3\rangle_t+\gamma\left(\frac{\hbar\omega}{2}-E_t\right).
\label{eq:IlluminatiHeatingEnergy}
\end{equation}
Using Eq.~\eqref{eq:defEt} and the rate Eq.~\eqref{eq:GammaIHO}, we obtain
\begin{equation}
\partial_t\langle\hat n\rangle_t=-\gamma\langle\hat n\rangle_t+8\Gamma_I^{\rm HO}\frac{\langle\hat K^3\rangle_t}{(\hbar\omega)^3}.
\label{eq:IlluminatiHeatingNumber}
\end{equation}
In contrast with the constant source of the Arzano model and the contribution proportional to $K_t$ in the Breuer model, the heating rate is thus governed by the third moment of the kinetic energy. Since $\avg{\hat K^3} \geq \avg{\hat K}^3$, the heating is strongly enhanced at large kinetic energies. As discussed below, taking a secular approximation predicts a finite-time divergence, although its validity is not guaranteed at high excitation.

Starting at $t=0$ from a Fock state $|n\rangle$, we have
\begin{equation}
\langle n|\hat K^3|n\rangle=\frac{(\hbar\omega)^3}{64}\left(20n^3+30n^2+40n+15\right),
\label{eq:IlluminatiK3Fock}
\end{equation}
so that the initial rate of change of the occupation is
\begin{equation}
\left.\partial_t\langle\hat n\rangle_t\right|_{t=0}=-\gamma n+\frac{\Gamma_I^{\rm HO}}{8}\left(20n^3+30n^2+40n+15\right).
\label{eq:IlluminatiInitialHeatingFock}
\end{equation}
For the ground state this gives the initial heating rate
\begin{equation}
\left.\partial_t\langle\hat n\rangle_t\right|_{t=0}=\frac{15}{8}\Gamma_I^{\rm HO},
\label{eq:IlluminatiInitialHeatingGround}
\end{equation}
for $|1\rangle$ it gives $-\gamma+105\Gamma_I^{\rm HO}/8$, and for $n\gg1$ the anomalous rate grows as $5\Gamma_I^{\rm HO}n^3/2$. The ground-state result is consistent with the depletion rate $6\Gamma_I^{\rm HO}/8$ of Eq.~\eqref{eq:IlluminatiHOpopulations}. The operator $\hat K^2\propto\hat p^4$ couples $|0\rangle$ to both $|2\rangle$ and $|4\rangle$, at the rates $9\Gamma_I^{\rm HO}/16$ and $3\Gamma_I^{\rm HO}/16$. Their sum gives the depletion rate, while weighting them by the number of phonons created gives $2\cdot\frac{9}{16}\Gamma_I^{\rm HO}+4\cdot\frac{3}{16}\Gamma_I^{\rm HO}=\frac{15}{8}\Gamma_I^{\rm HO}$, in agreement with Eq.~\eqref{eq:IlluminatiInitialHeatingGround}.

For states supported predominantly on Fock levels whose dissipative transition rates are small compared with $\omega$, a secular approximation extends Eq.~\eqref{eq:IlluminatiInitialHeatingFock} to the occupation dynamics. Averaging over an oscillation period removes the non-secular contributions, so that only the diagonal part of $\hat K^3$, given by Eq.~\eqref{eq:IlluminatiK3Fock}, contributes:
\begin{equation}
\partial_t\langle\hat n\rangle_t\simeq-\gamma\langle\hat n\rangle_t+\frac{\Gamma_I^{\rm HO}}{8}\left\langle f(\hat n)\right\rangle_t,\quad f(n)=20n^3+30n^2+40n+15,
\label{eq:IlluminatiSecularNumber}
\end{equation}
the analogue of Eq.~\eqref{eq:BreuerSecularNumber}. Since the secular transition rates grow as $\Gamma_I^{\rm HO}n^4$ at large $n$, a sufficient condition for this approximation is $\Gamma_I^{\rm HO}n^4\ll\omega$ over the significantly populated levels. This condition is necessarily violated upon extrapolating the secular dynamics to arbitrarily large occupation. 
A simple estimate follows from closing the system of equations of motion by taking $\langle f(\hat n)\rangle_t\simeq f(\langle\hat n\rangle_t)$, which is exact for Fock states and otherwise underestimates the heating rate, since $f$ is convex. 
Writing $f(n)=20(n+\tfrac12)^3+25(n+\tfrac12)$, Eq.~\eqref{eq:IlluminatiSecularNumber} without damping and for an initial ground state then gives
\begin{equation}
E_t=\hbar\omega\left[\frac{24}{5}e^{-25\Gamma_I^{\rm HO}t/4}-\frac45\right]^{-1/2},
\label{eq:IlluminatiClosure}
\end{equation}
which formally diverges at $t_I \equiv (4/25)\ln6/\Gamma_I^{\rm HO}\simeq0.29/\Gamma_I^{\rm HO}$ as $(t_I-t)^{-1/2}$, see Fig.~\ref{fig1}d. This singularity originates from the approximations taken above and does not necessarily establish a finite-time divergence of the full P--I master equation, since the secular approximation fails at sufficiently large occupation.

The closure, however, neglects the broadening of the phonon distribution. In the secular approximation, the populations obey a rate equation with transitions $n\to n\pm2$ and $n\to n\pm4$, whose rates grow as $n^4$. Formally extending these rates to arbitrarily high occupation yields a nonzero probability of reaching unbounded occupation in finite time. 
For an initial ground state without damping, this probability is approximately $1.83\times10^{-4}$ at $\Gamma_I^{\rm HO}t=0.01$ and $3.48\%$ at $\Gamma_I^{\rm HO}t=0.1$. In this idealized description, assigning infinite occupation to such trajectories makes the mean energy formally infinite for every $t>0$. Linear damping does not remove this behavior. However, since the secular approximation ceases to be valid when $\Gamma_I^{\rm HO}n^4$ becomes comparable to $\omega$, these results do not establish an energy divergence or exclude a stationary state in the full P--I dynamics. 
In the weak-noise regime $\Gamma_I^{\rm HO}\ll\gamma\ll\omega$, the low-occupation dynamics of an oscillator initially in its ground state admits an approximately quasistationary distribution, reached within a few damping times, with populations $P_1=3\Gamma_I^{\rm HO}/4\gamma$, $P_2=3\Gamma_I^{\rm HO}/8\gamma$, $P_3=\Gamma_I^{\rm HO}/16\gamma$ and $P_4=3\Gamma_I^{\rm HO}/64\gamma$ to first order in $\Gamma_I^{\rm HO}/\gamma$, hence $\langle\hat n\rangle\simeq15\Gamma_I^{\rm HO}/8\gamma$. Direct excitation of $|4\rangle$ and subsequent damping into $|3\rangle$ distinguish this distribution from those of the Breuer model and of thermal heating. This approximation applies over times long compared with $\gamma^{-1}$ but short compared with significant leakage into high-occupation states.

This hierarchy of state-dependent heating mechanisms can also arise in other quantum-gravity-inspired decoherence models. For instance, a Lindblad operator proportional to $\hat p^3$ would lead to an energy-injection term proportional to $\langle\hat K^2\rangle_t$~\cite{Lobo:2025mze}, an intermediate case between the Breuer and Petruzziello--Illuminati models. As in the latter, the dynamics would involve a hierarchy of higher moments and would not close at the level of the mean energy alone.

\subsection{Spectrum}
\label{sec:spectrum}

We now discuss the stationary position power spectral density, which provides a non-interferometric observable that can be accessed through continuous monitoring of the oscillator motion. Such a spectrum is well defined only when the dynamics approaches a steady state with finite second moments. For the Arzano model this is the case for any $\gamma>0$, whereas for the Breuer model we restrict to the stable regime Eq.~\eqref{eq:BreuerStabilityCondition}. For the Petruzziello--Illuminati model, we instead consider a leading-order quasistationary approximation.

We first consider the Arzano and Breuer models. In both cases, the dissipator depends only on $\hat p$ and therefore does not modify the adjoint equations for the first moments. Including amplitude damping, these read $\mathcal L^\dagger[\hat x]=\hat p/m-\gamma\hat x/2$ and $\mathcal L^\dagger[\hat p]=-m\omega^2\hat x-\gamma\hat p/2$. Consequently, for a time $s\geq0$, the position operator evolves as
\begin{equation}
\hat x(s)=e^{-\gamma s/2}\left[\hat x(0)\cos(\omega s)+\frac{\hat p(0)}{m\omega}\sin(\omega s)\right].
\label{eq:xAdjointEvolution}
\end{equation}

We use the quantum regression theorem \cite{breuer2002theory,gardiner2004quantum,guarnieri2014quantum} to evaluate the stationary two-time correlation function. Defining
\begin{equation}
C_{xx}^{\rm sym}(s)=\frac{1}{2}\langle\{\hat x(s),\hat x(0)\}\rangle_{\rm ss},
\end{equation}
Eq.~\eqref{eq:xAdjointEvolution} gives, for $s\geq0$,
\begin{equation}
C_{xx}^{\rm sym}(s)=e^{-\gamma s/2}\left[\langle\hat x^2\rangle_{\rm ss}\cos(\omega s)+\frac{S_{\rm ss}}{2m\omega}\sin(\omega s)\right],
\label{eq:CxxStationary}
\end{equation}
where $S_{\rm ss}=\langle\hat x\hat p+\hat p\hat x\rangle_{\rm ss}$, consistently with the notation of Sec.~\ref{sec:heating}. Stationarity implies $C_{xx}^{\rm sym}(-s)=C_{xx}^{\rm sym}(s)$.

The symmetrized position spectrum is defined as
\begin{equation}
S_{xx}(\mu)=\int_{-\infty}^{+\infty}\mathrm{d}s\,e^{i\mu s}C_{xx}^{\rm sym}(s).
\label{eq:SxxDefinition}
\end{equation}
Introducing the common denominator
\begin{equation}
\Lambda(\mu)=\left(\omega^2+\frac{\gamma^2}{4}-\mu^2\right)^2+\gamma^2\mu^2,
\label{eq:LambdaSpectrum}
\end{equation}
the Fourier transform of Eq.~\eqref{eq:CxxStationary} gives
\begin{equation}
S_{xx}(\mu)=\frac{\gamma\left(\omega^2+\mu^2+\frac{\gamma^2}{4}\right)\langle\hat x^2\rangle_{\rm ss}+\frac{S_{\rm ss}}{m}\left(\omega^2+\frac{\gamma^2}{4}-\mu^2\right)}{\Lambda(\mu)}.
\label{eq:SxxGeneral}
\end{equation}
Thus, the dynamical response entering the spectrum is the same for the Arzano and Breuer models. Their difference is encoded in the stationary second moments.

\textit{Arzano model.--} From Eq.~\eqref{eq:ArzanoMoments}, the stationary moments relevant to Eq.~\eqref{eq:SxxGeneral} are
\begin{equation}
\langle\hat x^2\rangle_{\rm ss}=\frac{\hbar}{2m\omega}+\frac{\hbar\Gamma_A^{\rm HO}}{m\omega}\frac{\gamma^2+2\omega^2}{\gamma(\gamma^2+4\omega^2)},\qquad S_{\rm ss}=-\frac{2\hbar\Gamma_A^{\rm HO}\omega}{\gamma^2+4\omega^2}.
\label{eq:ArzanoSpectrumMoments}
\end{equation}
Substitution into Eq.~\eqref{eq:SxxGeneral} yields
\begin{equation}
S_{xx}^{A}(\mu)=\frac{1}{\Lambda(\mu)}\left[\frac{\gamma\hbar}{2m\omega}\left(\omega^2+\mu^2+\frac{\gamma^2}{4}\right)+\frac{\hbar\Gamma_A^{\rm HO}}{m\omega}\left(\mu^2+\frac{\gamma^2}{4}\right)\right].
\label{eq:ArzanoSpectrum}
\end{equation}
The first term is the standard zero-temperature oscillator spectrum generated by the Hamiltonian dynamics and amplitude damping, while the second term is the excess noise induced by the Arzano dissipator. The latter is proportional to the same rate $\Gamma_A^{\rm HO}=m\omega c^2\tau_A/\hbar$ that controls the decoherence and heating discussed above.

\textit{Breuer model.--} In the stable regime, the stationary moments obtained from Eq.~\eqref{eq:BreuerMoments} are
\begin{subequations}\label{eq:BreuerSpectrumMoments}
\begin{align}
\langle\hat x^2\rangle_{\rm ss} &= \frac{\gamma\hbar(\gamma^2+4\omega^2+\gamma\Gamma_B^{\rm HO})}{2m\omega[\gamma(\gamma^2+4\omega^2)-2\Gamma_B^{\rm HO}\omega^2]},\\ 
S_{\rm ss} &= -\frac{\gamma\hbar\Gamma_B^{\rm HO}\omega}{\gamma(\gamma^2+4\omega^2)-2\Gamma_B^{\rm HO}\omega^2}.
\end{align}
\end{subequations}
The corresponding spectrum is
\begin{align}
S_{xx}^{B}(\mu) &= \frac{1}{\Lambda(\mu)}\left[\frac{\gamma\hbar}{2m\omega}\left(\omega^2+\mu^2+\frac{\gamma^2}{4}\right) + \right. \notag\\
& \left. \frac{\gamma\hbar\Gamma_B^{\rm HO}(\gamma^2+4\omega^2)}{2m\omega[\gamma(\gamma^2+4\omega^2)-2\Gamma_B^{\rm HO}\omega^2]}\left(\mu^2+\frac{\gamma^2}{4}\right)\right].
\label{eq:BreuerSpectrum}
\end{align}
As in the Arzano case, the first contribution is the ordinary zero-temperature oscillator spectrum and the second is the excess induced by the gravity-related dissipator. Well below the instability threshold, the latter is $\hbar\Gamma_B^{\rm HO}(\mu^2+\gamma^2/4)/[2m\omega\Lambda(\mu)]$. As the threshold $2\Gamma_B^{\rm HO}\omega^2=\gamma(\gamma^2+4\omega^2)$ is approached from below, the amplitude of the excess spectrum increases and diverges together with the stationary second moments.

Interestingly, Eqs.~\eqref{eq:ArzanoSpectrum} and \eqref{eq:BreuerSpectrum} show that the two models generate the same excess spectral lineshape,
\begin{equation}
S_{xx}^{\rm excess}(\mu)\propto\frac{\mu^2+\gamma^2/4}{\Lambda(\mu)},
\label{eq:ExcessSpectrumLineshape}
\end{equation}
despite their qualitatively different heating dynamics. At fixed oscillator parameters, the models therefore cannot be distinguished from the shape of the stationary position spectrum alone. Their spectral signatures differ instead through the amplitude and, in particular, through its dependence on the model parameters and on the oscillator frequency. Combining spectral measurements with the heating and decoherence observables discussed in the previous sections therefore provides additional discriminatory power.

\textit{Petruzziello--Illuminati model.--} The dissipator is again a function of $\hat p$ alone and leaves Eq.~\eqref{eq:xAdjointEvolution} unchanged, but the coherent correction $4a_P\overline{\beta}\hat K^2$ makes the evolution of $\hat x$ nonlinear in momentum, and the dissipator admits no stationary state with finite second moments, as shown in Sec.~\ref{sec:IlluminatiHeating}. 
In the experimentally relevant weak-noise regime $\gamma\gg\Gamma_I^{\rm HO}$, however, $\langle\hat K^3\rangle$ takes its ground-state value to leading order in $\Gamma_I^{\rm HO}/\gamma$. Neglecting the small coherent correction, Eqs.~\eqref{eq:IlluminatiMoments} then reduce to Eqs.~\eqref{eq:ArzanoMoments} with $\Gamma_A^{\rm HO}\to15\Gamma_I^{\rm HO}/4$. Thus, over observation times long compared with $\gamma^{-1}$ but short compared with significant high-occupation leakage, the leading-order quasistationary spectrum is approximated by Eq.~\eqref{eq:ArzanoSpectrum} with the same replacement. In this regime all three models produce the same excess lineshape, Eq.~\eqref{eq:ExcessSpectrumLineshape}.

The common lineshape nevertheless differs from those of conventional noise sources. A force noise, $-D_p[\hat x,[\hat x,\hat\rho]]$, equivalent to a classical white force with two-sided spectral density $2\hbar^2D_p$, adds the source $\hbar^2D_p/m$ to the equation for $\dot K_t$, while a thermal bath of mean occupation $n_{\rm th}$, obtained by replacing $\gamma\mathcal D[\hat a]$ with $\gamma(n_{\rm th}+1)\mathcal D[\hat a]+\gamma n_{\rm th}\mathcal D[\hat a^\dagger]$, adds $\gamma n_{\rm th}\hbar\omega/2$ to the equations for both $\dot K_t$ and $\dot U_t$. Neither modifies the first moments, and Eq.~\eqref{eq:SxxGeneral} gives the excess spectra
\begin{equation}
S_{xx}^{F}=\frac{2\hbar^2D_p}{m^2\Lambda(\mu)},\qquad S_{xx}^{\rm th}=\frac{\gamma\hbar n_{\rm th}}{m\omega}\,\frac{\omega^2+\mu^2+\gamma^2/4}{\Lambda(\mu)}.
\label{eq:conventionalSpectra}
\end{equation}
A force noise drives the momentum and gives a frequency-independent numerator, whereas the gravity-induced terms drive the position directly and give the numerator $\mu^2+\gamma^2/4$. The two lineshapes coincide near resonance, but normalized there their ratio is $(\mu^2+\gamma^2/4)/(\omega^2+\gamma^2/4)\simeq\mu^2/\omega^2$, so that the gravity-induced excess is suppressed below resonance and enhanced above it. Far above resonance, the gravity-induced terms fall off as $\mu^{-2}$, like a white velocity noise, e.g., $S_{xx}^A\to c^2\tau_A/\mu^2$ and $S_{xx}^B\to\tau_B\langle\hat p^2\rangle_{\rm ss}/m^2\mu^2$, whereas a force noise falls off as $\mu^{-4}$. Broadband displacement spectra therefore break this degeneracy, and likewise distinguish the gravity-induced terms from spring-constant noise, which acts as a force-like source. The thermal excess also falls off as $\mu^{-2}$, but this reflects the amplitude-damping model. For viscous (Caldeira--Leggett) damping, thermal noise is a pure force noise~\cite{breuer2002theory}, so a quantitative analysis of the wings requires a damping model valid away from resonance.

\subsection{Classical-noise interpretation}
\label{sec:classicalNoise}

The heating mechanisms derived above have a clear interpretation in terms of the stochastic unraveling already used for the free particle. Writing the gravity-induced dissipators as $-\frac12[\hat C_X,[\hat C_X,\hat\rho]]$, as in Eq.~\eqref{eq:eomOpO}, the collapse operators are
\begin{equation}
\hat C_A=\frac{c\sqrt{\tau_A}}{\hbar}\,\hat p, \qquad \hat C_B=\frac{\sqrt{\tau_B}}{\hbar}\,\hat K, \qquad \hat C_I=\frac{4a_P\sqrt{\tau_I}}{\hbar}\,\hat K^2,
\end{equation}
with $X=A,B,I$ labeling the model. Since they are Hermitian, the master equations~\eqref{eq:ArzanoModel}, \eqref{eq:BreuerModel} and \eqref{eq:IlluminatiModel} are ensemble averages, $\hat\rho_t=\mathbb E[|\psi_t\rangle\langle\psi_t|]$, of the stochastic Schr\"odinger equation~\cite{arnold1974stochastic}
\begin{equation}
i\hbar\frac{d}{dt}|\psi_t\rangle = \left[\hat H_X+\hbar\hat C_X\xi(t)\right]|\psi_t\rangle,
\label{eq:classical_noise_SSE}
\end{equation}
understood in the Stratonovich sense, with $\xi(t)$ the real Gaussian white noise satisfying $\mathbb E[\xi(t)]=0$ and $\mathbb E[\xi(t)\xi(t')]=\delta(t-t')$. Here $\hat H_A=\hat H_B=\hat H_0$, the harmonic oscillator Hamiltonian Eq.~\eqref{eq:HHOdef}, and $\hat H_I=\hat H_0+4a_P\overline{\beta}\hat K^2$. Each realization evolves unitarily, and since $\hat C_X=f_X(\hat p)$, the Heisenberg equations read
\begin{equation}
\dot{\hat x}=\frac{i}{\hbar}[\hat H_X,\hat x]+\hbar f_X'(\hat p)\,\xi(t),\qquad \dot{\hat p}=-m\omega^2\hat x,
\label{eq:classicalNoiseHeisenberg}
\end{equation}
with $\hbar f_A'=c\sqrt{\tau_A}$, $\hbar f_B'=\sqrt{\tau_B}\,\hat p/m$ and $\hbar f_I'=4a_P\sqrt{\tau_I}\,\hat p^3/m^2$. As for the free particle, the noise enters only the velocity, and the resulting position diffusion now heats the oscillator at rate
\begin{equation}
\left.\frac{dE_t}{dt}\right|_{\mathrm{noise}} = \frac{m\omega^2\hbar^2}{2} \bigl\langle[f_X'(\hat p)]^2\bigr\rangle_t.
\label{eq:classical_noise_heating}
\end{equation}

For the Arzano model, the velocity noise is additive and independent of the state, and Eq.~\eqref{eq:classical_noise_heating} gives the constant rate $\hbar\omega\Gamma_A^{\rm HO}/2$ of Eq.~\eqref{eq:ArzanoHeatingEnergy}. For the Breuer model, the velocity noise is proportional to the velocity itself, i.e., the noise modulates the coefficient of the kinetic term, and Eq.~\eqref{eq:classical_noise_heating} gives $\Gamma_B^{\rm HO}\langle\hat K\rangle_t$. This is a parametric heating process producing the exponential growth discussed in Sec.~\ref{sec:BreuerHeating}. For $\Gamma_B^{\rm HO}\ll\omega$, averaging over the oscillation gives $\langle\hat K\rangle_t\simeq E_t/2$, and thus
\begin{equation}
\frac{dE_t}{dt}\simeq\Gamma_{\rm par}E_t, \qquad \Gamma_{\rm par} =\frac{\Gamma_B^{\mathrm{HO}}}{2} =\frac{\omega^2\tau_B}{2},
\label{eq:BreuerParametric}
\end{equation}
which coincides with the standard parametric-heating rate $\Gamma_{\rm par}=\pi^2\nu^2S_\epsilon(2\nu)$~\cite{Savard1997}, with $\nu=\omega/2\pi$ and $S_\epsilon=2\tau_B$ the one-sided spectrum of $\epsilon(t)=\sqrt{\tau_B}\xi(t)$. Reference~\cite{Savard1997} considers fluctuations of the potential term, but exchanging the position and momentum quadratures gives the same rate for fluctuations of the kinetic term. 
For the Petruzziello--Illuminati model, the velocity noise grows as $\hat p^3$, and Eq.~\eqref{eq:classical_noise_heating} gives the term $64a_P^2\tau_I\omega^2\langle\hat K^3\rangle_t$ of Eq.~\eqref{eq:IlluminatiEnergyFull}, whereas the coherent term there originates from the deterministic correction $4a_P\overline\beta\hat p^3/m^2$ to the velocity and does not represent irreversible heating.

The three heating mechanisms thus share a common origin, a velocity noise with momentum dependence proportional to $1$, $\hat p$ and $\hat p^3$, which yields energy-injection rates that are constant, proportional to $\langle\hat K\rangle_t$ and proportional to $\langle\hat K^3\rangle_t$, respectively. This interpretation is connected with the emergence of Lindblad dynamics from the quantization of stochastic modified dispersion relations~\cite{Lobo:2025mze}.

\section{Experimental bounds}
\label{sec:experimentalHeating}

The anomalous heating predicted by the three models provides a direct route to experimental constraints. Suppose that, after accounting for known environmental and technical noise sources, an experiment places an upper bound $R_{\rm exp}$ on an additional phonon-heating rate. For a single particle of mass $m$ in a harmonic potential, initially in the ground state, Eqs.~\eqref{eq:ArzanoHeatingShort}, \eqref{eq:BreuerInitialHeating} and \eqref{eq:IlluminatiInitialHeatingGround} give the anomalous heating rates
\begin{subequations}\label{eq:experimentalHeatingRates}
\begin{align}
& R_A=\frac{\Gamma_A^{\rm HO}}{2}=\frac{m\omega c^2}{2\hbar}\tau_A,\qquad R_B=\frac{\Gamma_B^{\rm HO}}{4}=\frac{\omega^2}{4}\tau_B,\\ 
&\qquad\qquad R_I=\frac{15\Gamma_I^{\rm HO}}{8}=15\frac{m^2\ell_P^4\omega^4}{\hbar^2}\tau_I,
\end{align}
\end{subequations}
so that a measured upper limit $R_{\rm exp}$ implies
\begin{equation}
\tau_A\lesssim\frac{2\hbar R_{\rm exp}}{m\omega c^2},\qquad \tau_B\lesssim\frac{4R_{\rm exp}}{\omega^2},\qquad \tau_I\lesssim\frac{\hbar^2R_{\rm exp}}{15m^2\ell_P^4\omega^4}.
\label{eq:experimentalTauBounds}
\end{equation}
At fixed $R_{\rm exp}$, large oscillator frequencies thus improve the sensitivity to all three models. Excited states provide a further handle on the latter two, since for an initial Fock state $|n\rangle$ the Breuer contribution grows as $R_B(n)=\Gamma_B^{\rm HO}(2n+1)/4$, while the Petruzziello--Illuminati contribution, $R_I(n)=\Gamma_I^{\rm HO}(20n^3+30n^2+40n+15)/8$, scales as $n^3$ for $n\gg1$.

\subsection{Composite oscillators}
\label{sec:compositeModes}

Real mechanical oscillators are composite objects made of many atoms, and the extension of the single-particle master equations~\eqref{eq:ArzanoModel}--\eqref{eq:IlluminatiModel} to such systems is neither unique nor straightforward. 
One must specify which constituents are fundamental, how the gravity-related fluctuations act on them, and how these fluctuations are correlated between different constituents.
For the Arzano model, one must also specify how its deformed composition law of momenta applies to a composite body, an issue also known as the soccer-ball problem~\cite{amelino-camelia_challenge_2013,Hossenfelder2014}. In Appendix~\ref{sec:compositeMain}, we apply the single-particle dissipators to each of the $N$ atoms of mass $m=M/N$, with a matrix $C_{ab}$, normalized as $C_{aa}=1$, describing the correlations between the fluctuations acting on atoms $a$ and $b$. For the rigid center-of-mass motion, and neglecting the internal motion of the atoms, the characteristic oscillator rates Eqs.~\eqref{eq:HOdecoherenceRates} become
\begin{subequations}\label{eq:CMOscillatorRates}
\begin{align}
&\Gamma_{A,\mathrm{CM}}^{\mathrm{HO}}=\frac{M\omega c^2\tau_A}{\hbar}\frac{\mathcal C}{N^2},\qquad \Gamma_{B,\mathrm{CM}}^{\mathrm{HO}}=\tau_B\omega^2\frac{\mathcal C}{N^2},\\
&\quad\qquad \Gamma_{I,\mathrm{CM}}^{\mathrm{HO}}=8(a_P^{\mathrm{CM}})^2\tau_I\hbar^2\omega^4\frac{\mathcal C}{N^6},
\end{align}
\end{subequations}
with $a_P^{\rm CM}=M\ell_P^2/\hbar^2$ and $\mathcal C=\sum_{a,b}C_{ab}$. The latter ranges from $\mathcal C=N$ for independent fluctuations (i.e. $C_{ab}=\delta_{ab}$), to $\mathcal C=N^2$ for fully collective fluctuations (i.e. $C_{ab}=1$). 
For statistically homogeneous fluctuations, $C_{ab}=C(\mathbf r_a-\mathbf r_b)$ decays over a correlation length $\ell_c$, and the two limits correspond to $\ell_c$ much shorter than the interatomic distance and much longer than the size of the body, respectively.
The single-particle rates with $m$ replaced by $M$ are thus recovered for the Arzano and Breuer models only for collective fluctuations, whereas independent fluctuations reduce both rates by a factor $N$. For the Petruzziello--Illuminati model, even collective fluctuations reduce the rate by $N^4$. The corresponding heating rates and bounds follow from Eqs.~\eqref{eq:experimentalHeatingRates} and \eqref{eq:experimentalTauBounds} with these replacements.
A measured upper limit $R_{\rm exp}$ on the center-of-mass heating rate then implies
\begin{equation}
\tau_A\lesssim\frac{2\hbar R_{\rm exp}}{M\omega c^2}\frac{N^2}{\mathcal C},\quad \tau_B\lesssim\frac{4R_{\rm exp}}{\omega^2}\frac{N^2}{\mathcal C},\quad \tau_I\lesssim\frac{\hbar^2R_{\rm exp}}{15M^2\ell_P^4\omega^4}\frac{N^6}{\mathcal C},
\label{eq:experimentalTauCMBounds}
\end{equation}
i.e., Eq.~\eqref{eq:experimentalTauBounds} with $m\to M$, weakened by the factor $N^2/\mathcal C$ for the Arzano and Breuer models and by $N^6/\mathcal C$ for the Petruzziello--Illuminati model.
Neglecting the internal motion is exact for the Arzano model and for collective Breuer fluctuations. Otherwise, the thermal and zero-point motion of the atoms adds heating rate contributions that are not suppressed by $N$, as discussed in Appendix~\ref{sec:compositeMain}.

\subsection{Separating gravity-induced from conventional heating}

The bounds discussed above conservatively attribute the entire observed heating to a gravity-related mechanism, but three features of the models allow one to go beyond this. First, conventional heating from photon recoil or force noise does not depend on the oscillator excitation, whereas the Breuer and Petruzziello--Illuminati contributions grow as $n$ and $n^3$, respectively. If the measured rate $R_{\rm exp}(n)$ remains independent of $n$, the Breuer bound improves as $\tau_B\lesssim4R_{\rm exp}(n)/[(2n+1)\omega^2]$. This criterion does not, however, exclude parametric technical noise, such as fluctuations of the laser intensity or of the trap frequency, whose heating rate also grows as $n$~\cite{Savard1997}, since the Breuer term is itself a parametric noise, Eq.~\eqref{eq:BreuerParametric}. Similarly, fluctuations of the trap anharmonicity can produce heating growing as $n^3$. Such technical contributions can be identified through their dependence on the trap power and on the laser noise. Second, the spectral wings discussed in Sec.~\ref{sec:spectrum} distinguish gravity-induced noise from force and spring-constant noise, although not the three models from each other. Third, the gravity-induced rates are independent of the trapping mechanism and follow the frequency, mass, and correlation dependences of Eqs.~\eqref{eq:experimentalHeatingRates} and \eqref{eq:CMOscillatorRates}, which can be tested across modes of the same device or across platforms.

\subsection{Experimental bounds on the Arzano model}

We evaluate the bounds on $\tau_A$ in two interpretations. In the single-particle interpretation, the oscillator is treated as an elementary particle of mass $M$, and Eq.~\eqref{eq:experimentalTauBounds} applies with $m\to M$. In the composite interpretation, the bound depends on the correlations through Eq.~\eqref{eq:experimentalTauCMBounds}, between the collective case, $\mathcal C=N^2$, and the independent case, $\mathcal C=N$. For the center of mass, the single-particle interpretation coincides with collective fluctuations, and for the Arzano model neglecting the internal motion is exact. 
Every position readout, moreover, measures the displacement of the oscillator relative to a reference, meaning that fluctuations correlated between oscillator and reference cancel in the readout. The collective case thus requires fluctuations correlated within the oscillator but not with the reference, i.e., $\ell_c$ larger than the oscillator but shorter than its distance to the reference.

\textit{Center-of-mass motion.--} For collective fluctuations, the velocity noise $c\sqrt{\tau_A}\,\xi(t)$ of Eq.~\eqref{eq:classicalNoiseHeisenberg} is independent of mass, so that the most sensitive detectors are those with the lowest displacement noise, namely free test masses. 
These have $\omega,\gamma\to0$, such that Eq.~\eqref{eq:ArzanoSpectrum} gives the one-sided position spectral density $S_{xx}=2c^2\tau_A/\mu^2$ or, through double differentiation, the acceleration spectral density $S_{a}=\mu^4 S_{xx}$.
Since experiments read out the differential motion of separate free test masses, assuming independent velocity noise on each mass requires $\ell_c$ to be shorter than their separation, about $38\,\mathrm{cm}$ for LISA Pathfinder and $4\,\mathrm{km}$ for Advanced LIGO. 
This gives $\tau_A\lesssim S_a/(4\mu^2c^2)$ for the two test masses of LISA Pathfinder and $\tau_A\lesssim\mu^2S_{\Delta L}/(8c^2)$ for the four of Advanced LIGO, with $S_a$ and $S_{\Delta L}$ the measured acceleration and differential-displacement noise. 
The LISA Pathfinder noise, $1.74\,\mathrm{fm\,s^{-2}/\sqrt{\rm Hz}}$ above $2\,\mathrm{mHz}$~\cite{Armano2018}, gives $\tau_A\lesssim2\times10^{-45}\,\mathrm{s}\simeq0.04\,t_P$ at $10\,\mathrm{mHz}$, which would tighten at the highest frequency where $S_a$ remains flat. 
The Advanced LIGO strain noise, below $10^{-23}\,\mathrm{Hz^{-1/2}}$ near $100\,\mathrm{Hz}$~\cite{Buikema2020}, gives $\tau_A\lesssim10^{-51}\,\mathrm{s}$, nearly eight orders of magnitude below the Planck time. 
The levitated nanoparticle of Ref.~\cite{Piotrowski2023} gives $\tau_A\lesssim2.6\times10^{-36}\,\mathrm{s}$, about $5\times10^7$ times the Planck time. 
For fluctuations independent also between constituent atoms, all these bounds weaken by $N\simeq M/m$, to $\tau_A\lesssim10^{-20}\,\mathrm{s}$ for LISA Pathfinder, $10^{-24}\,\mathrm{s}$ for Advanced LIGO, and $3\times10^{-28}\,\mathrm{s}$ for the nanoparticle. As for collapse models~\cite{Carlesso2016,Helou2017}, these bounds rely only on displacement noise, not on superposition states.

\textit{Internal acoustic modes.--} Collective fluctuations kick all atoms identically and thus displace the body rigidly, without exciting its internal modes. 
For an internal mode, homogeneous correlations rescale the single-particle mass $m$ to an effective mass $m_A$, which is largest, $m_A\sim\varrho(\lambda/2)^3$, when $\ell_c\sim\lambda/2$ matches the mode shape, and reduces to the atomic mass for fluctuations independent between atoms (Appendix~\ref{sec:compositeMain}). 
The Arzano term adds $\Gamma_A^{\rm HO}/2\gamma$ to the stationary occupation, Eq.~\eqref{eq:ArzanoHeatingNumber}, independently of the bath temperature. 
Comparing the measured occupation with the one expected from the bath therefore gives $\tau_A\lesssim2\epsilon k_BT/(Q\,m_A\omega c^2)$, with $Q=\omega/\gamma$, for a thermal mode whose occupation agrees with $n_{\rm th}\simeq k_BT/\hbar\omega$ to a relative precision $\epsilon\equiv|\langle\hat n\rangle_\infty-n_{\rm th}|/n_{\rm th}$. 
For a mode close to its ground state the bound becomes $\tau_A\lesssim2\gamma n_{\rm ex}\hbar/(m_A\omega c^2)$, with $n_{\rm ex}$ the residual occupation. 
Unlike the Breuer excess discussed below, this temperature-independent excess appears as an additive offset in $\langle\hat n\rangle_\infty(T)$, separable from a calibration error e.g. by sweeping the bath temperature~\cite{Vinante2016,omahen_ultracold_2026}. For $\ell_c\sim\lambda/2$ and $\epsilon=0.1$, a $200\,\mathrm{MHz}$ mode of the quartz BAW resonator~\cite{Galliou2013}, with $\lambda/2\simeq16\,\mu\mathrm{m}$, gives $\tau_A\lesssim10^{-48}\,\mathrm{s}$, and for $n_{\rm ex}=10^{-2}$ the silicon nanobeam~\cite{MacCabe2020} and the HBAR~\cite{Bild2023} give $2\times10^{-48}\,\mathrm{s}$ and $3\times10^{-45}\,\mathrm{s}$, respectively. 
For the nanobeam, whose cross-section is smaller than $\lambda/2\simeq0.8\,\mu\mathrm{m}$, we take $m_A\simeq2\times10^{-16}\,\mathrm{kg}$, the mass of a beam segment of length $\lambda/2$. 
Correlations extending over the full beam width reduce $m_A$, and weaken this bound, by up to an order of magnitude, since the breathing mode displaces the two halves of the beam in opposite directions. The longitudinal overtones of a multimode BAW resonator span $\lambda/2\simeq4$--$200\,\mu\mathrm{m}$, so that one device can scan $\ell_c$ over almost two orders of magnitude. For independent fluctuations, the nanobeam gives $\tau_A\lesssim10^{-38}\,\mathrm{s}$, and the HBAR and quartz BAW resonators $\tau_A\lesssim(2\text{--}4)\times10^{-34}\,\mathrm{s}$. 
For $\ell_c\lesssim1\,\mu\mathrm{m}$, including independent fluctuations, bulk heating gives tighter bounds, whereas acoustic modes remain more sensitive at longer $\ell_c$.

\subsection{Experimental bounds on the Breuer model}
\label{sec:BreuerFDT}

Since $\Gamma_B^{\rm HO}=\tau_B\omega^2$ is independent of mass and, for collective fluctuations, unchanged for any mode of a composite body (Appendix~\ref{sec:compositeMain}), the single-particle interpretation and the composite interpretation with collective fluctuations predict the same rate, and the bounds below depend on the oscillator only through its frequency, quality factor and occupation. 
For independent fluctuations, the direct rate is multiplied by the inverse number of atoms participating in the mode, but the thermal and zero-point motion of the atoms, characterized by a kinetic temperature $T_{\rm kin}$, adds the state-independent rate $\Gamma_{B,\rm eff}=\omega\tau_Bk_BT_{\rm kin}/\hbar$. 
This makes the bounds tighter than for collective fluctuations, especially at low temperature, where zero-point motion keeps $T_{\rm kin}$ well above the bath temperature.

Since all platforms considered here reach a finite stationary occupation, they operate below the instability threshold, which already implies $\Gamma_B^{\rm HO}<2\gamma$. With a thermal bath of occupation $n_{\rm th}$, Eq.~\eqref{eq:BreuerSecularNumber} becomes
\begin{equation}
\partial_t\langle\hat n\rangle_t\simeq-\gamma_{\rm meas}\langle\hat n\rangle_t+\gamma n_{\rm th}+\frac{\Gamma_B^{\rm HO}}{4},\qquad \gamma_{\rm meas}\equiv\gamma-\frac{\Gamma_B^{\rm HO}}{2},
\label{eq:BreuerSecularThermal}
\end{equation}
where $\gamma_{\rm meas}$, the relaxation rate of $\langle\hat n\rangle_t$, differs from the linewidth $\gamma$ by an amount irrelevant for the order of magnitude of the bounds. The stationary occupation,
\begin{equation}
\langle\hat n\rangle_\infty=n_{\rm th}+\frac{\Gamma_B^{\rm HO}(2n_{\rm th}+1)}{4\gamma_{\rm meas}},
\label{eq:BreuerThermal}
\end{equation}
exceeds $n_{\rm th}\simeq k_BT/\hbar\omega$, and comparing the two as for the Arzano model gives for a mode in a thermal state
\begin{equation}
\tau_B\lesssim\frac{2\epsilon}{Q\,\omega},\qquad Q\equiv\frac{\omega}{\gamma_{\rm meas}},
\label{eq:BreuerFDT}
\end{equation}
and for a mode near the ground state $\tau_B\lesssim4\gamma_{\rm meas}n_{\rm ex}/\omega^2$, consistently with Eq.~\eqref{eq:BreuerStationaryWeak}. Unlike the Arzano excess, the Breuer excess is a fixed fraction of $n_{\rm th}$ for $n_{\rm th}\gg1$, degenerate with a calibration error, so that $\epsilon$ must be obtained from an absolute calibration of the readout, for instance with a known applied force. Any damping not originating from the bath, such as dynamical backaction or feedback, must likewise be calibrated.

\textit{Center-of-mass motion.--} For free test masses at rest, the direct Breuer position diffusion, $\tau_B\langle\hat p^2\rangle/M^2$, is negligible. For independent fluctuations, however, the internal motion acts as a velocity noise of strength $\tau_Bk_BT_{\rm kin}/M$, so that the free-mass bounds on the Arzano model translate into $\tau_B\lesssim8\times10^{-13}\,\mathrm{s}$ for Advanced LIGO and $10^{-7}\,\mathrm{s}$ for LISA Pathfinder.
For the levitated nanoparticle of Ref.~\cite{Piotrowski2023}, the ground-state heating rate gives $\tau_B\lesssim4R_{\rm exp}/\omega^2\simeq9\times10^{-9}\,\mathrm{s}$ in the single-particle interpretation, and $\Gamma_{B,\rm eff}/2\leq R_{\rm exp}$ gives $\tau_B\lesssim2\hbar R_{\rm exp}/(\omega k_BT_{\rm kin})\simeq6\times10^{-17}\,\mathrm{s}$ for independent fluctuations, taking $T_{\rm kin}\simeq1000$~K, as measured for a laser-trapped silica nanoparticle~\cite{Hebestreit2018}.

\textit{Internal acoustic modes.--} Since the bounds depend on the oscillator only through its frequency, quality factor and occupation, the figure of merit of a thermal mode is the $Q\times f$ product, which favors ultracoherent resonators. For $\epsilon=0.1$, Eq.~\eqref{eq:BreuerFDT} gives $\tau_B\lesssim3\times10^{-17}\,\mathrm{s}$ for soft-clamped~\cite{Tsaturyan2017} $\mathrm{Si_3N_4}$ strings with $Q\simeq8\times10^8$ at $1.3\,\mathrm{MHz}$ and room temperature~\cite{Ghadimi2018}, and $\tau_B\lesssim2\times10^{-20}\,\mathrm{s}$ for phonon-trapping quartz BAW resonators with $Q\times f\simeq1.6\times10^{18}\,\mathrm{Hz}$ at $4\,\mathrm{K}$~\cite{Galliou2013}. For ground-state gigahertz modes with $n_{\rm ex}=10^{-2}$, the HBAR of Ref.~\cite{Bild2023}, at $6\,\mathrm{GHz}$ with $T_1\simeq84\,\mu\mathrm{s}$, gives $\tau_B\lesssim3\times10^{-19}\,\mathrm{s}$, compared with the measured bound $\tau_B\lesssim4\times10^{-18}\,\mathrm{s}$ from the decay of its phonon Fock and superposition states~\cite{donadi_quantum_2025}. The silicon nanobeam of Ref.~\cite{MacCabe2020}, at $5\,\mathrm{GHz}$ with $T_1\simeq1.5\,\mathrm{s}$, gives $\tau_B\lesssim3\times10^{-23}\,\mathrm{s}$, fourteen orders of magnitude below the levitated nanoparticle due to a $\sim 10^4$ higher frequency (which enters squared) and a $\sim 10^6$ times lower residual heating rate $\gamma_{\rm meas}n_{\rm ex}\simeq7\times10^{-3}\,\mathrm{s^{-1}}$.
For independent fluctuations, $\Gamma_{B,\rm eff}$ adds the fractional excess $(\tau_B\omega Q/2)(T_{\rm kin}/T)$, which tightens Eq.~\eqref{eq:BreuerFDT} by $T_{\rm kin}/T\geq1$, up to $\sim10^2$ for cryogenic resonators, and gives $\tau_B\lesssim2\gamma_{\rm meas}n_{\rm ex}\hbar/(\omega k_BT_{\rm kin})$ for ground-state modes, i.e., $10^{-26}\,\mathrm{s}$ for the nanobeam and $10^{-22}\,\mathrm{s}$ for the HBAR, with $T_{\rm kin}\simeq240$~K and $390$~K set by zero-point motion even at environment temperatures of a few millikelvin.

\subsection{Experimental bounds on the Petruzziello--Illuminati model}

Writing $\beta(t)=\overline{\beta}+\delta\beta(t)$ with $\langle\delta\beta(t)\delta\beta(t')\rangle=\sigma_\beta^2\tau_c\,\delta(t-t')$ shows that Eq.~\eqref{eq:IlluminatiModel} corresponds to $\tau_I=\sigma_\beta^2\tau_c$, so that the natural benchmark is $\sigma_\beta^2t_P$ rather than the Planck time itself. In the single-particle interpretation, the oscillator is treated as an elementary particle of mass $M$, or of the effective mass of the mode. In the composite interpretation, the direct term is suppressed by $N^6/\mathcal C$, i.e., by at least $N^4$, but the atomic motion adds the state-independent rate $\Gamma_{I,\rm eff}\propto\omega(k_BT_{\rm kin})^3$, which involves only the atomic mass, is independent of $N$, and dominates for both independent and collective fluctuations (Appendix~\ref{sec:compositeMain}). We use its value for independent fluctuations, which for the center of mass differs from the collective one only by a factor $3.5$.

In the single-particle interpretation, ground-state heating is not the optimal strategy. For a thermal state of mean occupation $\bar n$, Gaussian moments give $\langle\hat K^3\rangle=15(\hbar\omega)^3(2\bar n+1)^3/64$, and Eq.~\eqref{eq:IlluminatiHeatingNumber} yields the anomalous heating rate $\frac{15}{8}\Gamma_I^{\rm HO}(2\bar n+1)^3$, which reduces to Eq.~\eqref{eq:IlluminatiInitialHeatingGround} for $\bar n=0$. A fluctuation--dissipation test of a mode with $n_{\rm th}\gg1$, with $\epsilon$ defined as for the Arzano model, then implies, to leading order in $\Gamma_I^{\rm HO}$,
\begin{equation}
\tau_I\lesssim\frac{\epsilon\hbar^4}{120\,Q\,M^2\ell_P^4\,\omega\,k_B^2T^2},
\label{eq:IlluminatiFDT}
\end{equation}
favoring massive, high-$Q$ modes at high temperature. 
This bound is a leading-order weak-noise estimate based on the thermal-state heating rate, and its interpretation as a steady-state excess occupation requires the dynamics to remain close to the assumed thermal state over the measurement time.
In the composite interpretation, this advantage disappears, since $\Gamma_{I,\rm eff}$ depends only on $T_{\rm kin}$ and not on the state of the mode.

\textit{Center-of-mass motion.--} For the levitated nanoparticle of Ref.~\cite{Piotrowski2023}, the single-particle interpretation gives $\tau_I\lesssim\hbar^2R_{\rm exp}/(15M^2\ell_P^4\omega^4)\simeq7\times10^{83}\,\mathrm{s}$, i.e., $\sigma_\beta\lesssim4\times10^{63}$ for $\tau_c=t_P$. In the composite interpretation, $\Gamma_{I,\rm eff}/2\leq R_{\rm exp}$ gives the tighter bound $\tau_I\lesssim10^{75}\,\mathrm{s}$, i.e., $\sigma_\beta\lesssim10^{59}$, for $T_{\rm kin}\simeq1000$~K~\cite{Hebestreit2018}, a bound scaling as $T_{\rm kin}^{-3}$.

\textit{Internal acoustic modes.--} In the single-particle interpretation, Eq.~\eqref{eq:IlluminatiFDT} applied to the quartz BAW resonator, with an effective mass of order $0.1\,\mathrm{mg}$~\cite{Galliou2013}, gives $\tau_I\lesssim5\times10^{39}\,\mathrm{s}$, i.e., $\sigma_\beta\lesssim3\times10^{41}$, for $\epsilon=0.1$. In the composite interpretation, the best acoustic mode is the silicon nanobeam, which for $n_{\rm ex}=10^{-2}$ gives $\tau_I\lesssim2\gamma n_{\rm ex}\hbar/[560\,a_P^2\omega(k_BT_{\rm kin})^3]\simeq2\times10^{66}\,\mathrm{s}$, whereas the HBAR, the quartz BAW and the $\mathrm{Si_3N_4}$ string give $10^{70}$ to $10^{75}\,\mathrm{s}$. Bulk heating, which sums the same internal-motion contribution over all the modes of a solid, gives the much tighter bound $\tau_I\lesssim10^{49}\,\mathrm{s}$ (Sec.~\ref{sec:bulkHeating}).
For comparison, the decay of phonon Fock and superposition states of the HBAR, treated as an elementary oscillator of effective mass $16.2\,\mu\mathrm{g}$, gives $\tau_I\lesssim4\times10^{46}\,\mathrm{s}$~\cite{donadi_quantum_2025}.

\textit{Coherent correction.--} The same platforms can constrain the coherent correction $4a_P\overline{\beta}\hat K^2$ far more tightly. After the rotating-wave approximation, it produces an amplitude-dependent frequency shift, whose measurement with micromechanical oscillators~\cite{Bawaj:2014cda}, and with a sapphire split-bar resonator and quartz BAW resonators~\cite{Bushev2019}, bounds the static GUP parameter at the level of $5\times10^6$, up to normalization conventions. 
Heating nonetheless offers an independent fingerprint, the $n^3$ scaling of the direct term, which could be mimicked only by fluctuations of the trap anharmonicity, although with a different dependence on the trapping parameters.

\begin{table*}[t!]
\caption{\textbf{Bounds on the three models from mechanical platforms and bulk heating.} Illustrative bounds (in seconds) on $\tau_A$, $\tau_B$ and $\tau_I$ in the single-particle interpretation, equivalent to collective fluctuations for $\tau_A$ and $\tau_B$, and in the composite interpretation with independent fluctuations, including the internal-motion contributions of Eqs.~(\ref{eq:BreuerCrossTerm}, \ref{eq:IlluminatiCrossTerm}). 
The Planck time is $t_P\simeq5.4\times10^{-44}\,\mathrm{s}$. Values for the free masses, the levitated nanoparticle and bulk heating use measured noise or heating levels. Values for the other platforms assume a calibration precision $\epsilon=0.1$ for the $\mathrm{Si_3N_4}$ string and the quartz BAW resonator, and a residual occupation $n_{\rm ex}=10^{-2}$ for the ground-state gigahertz modes.}
\label{tab:bounds}
\begin{ruledtabular}
\begin{tabular}{llcccccc}
& & \multicolumn{2}{c}{Arzano, $\tau_A$} & \multicolumn{2}{c}{Breuer, $\tau_B$} & \multicolumn{2}{c}{Petruzziello--Illuminati, $\tau_I$}\\
\cline{3-4}\cline{5-6}\cline{7-8}
Platform & Information & correlated & independent & collective & independent & elementary & composite\\
\hline
Advanced LIGO~\cite{Buikema2020} & $100\,\mathrm{Hz}$, $h\lesssim10^{-23}\,\mathrm{Hz^{-1/2}}$ & $10^{-51}$\,$^{a}$ & $10^{-24}$ & --\,$^{c}$ & $8\times10^{-13}$\,$^{d}$ & -- & --\\
LISA Pathfinder~\cite{Armano2018} & $10\,\mathrm{mHz}$, $1.74\,\mathrm{fm\,s^{-2}/\sqrt{Hz}}$ & $2\times10^{-45}$\,$^{a}$ & $10^{-20}$ & --\,$^{c}$ & $10^{-7}$\,$^{d}$ & -- & --\\
Levitated nanoparticle~\cite{Piotrowski2023} & $268\,\mathrm{kHz}$, $R_{\rm exp}=2\pi\times1\,\mathrm{kHz}$ & $3\times10^{-36}$\,$^{a}$ & $3\times10^{-28}$ & $9\times10^{-9}$ & $6\times10^{-17}$\,$^{e}$ & $7\times10^{83}$ & $10^{75}$\,$^{e}$\\
$\mathrm{Si_3N_4}$ string~\cite{Ghadimi2018} & $1.3\,\mathrm{MHz}$, $Q=8\times10^8$, $300\,\mathrm{K}$ & -- & -- & $3\times10^{-17}$ & $3\times10^{-17}$ & -- & $3\times10^{75}$\\
Quartz BAW~\cite{Galliou2013} & $200\,\mathrm{MHz}$, $Q=8\times10^{9}$, $4\,\mathrm{K}$ & $10^{-48}$\,$^{b}$ & $4\times10^{-34}$ & $2\times10^{-20}$ & $4\times10^{-22}$\,$^{f}$ & $5\times10^{39}$ & $2\times10^{71}$\,$^{f}$\\
HBAR~\cite{Bild2023} & $6\,\mathrm{GHz}$, $T_1=84\,\mu\mathrm{s}$ & $3\times10^{-45}$\,$^{b}$ & $2\times10^{-34}$ & \makecell{$3\times10^{-19}$\\ $4\times10^{-18}$\,$^{i}$} & $10^{-22}$\,$^{f}$ & $4\times10^{46}$\,$^{i}$ & $10^{70}$\,$^{f}$\\
Si nanobeam~\cite{MacCabe2020} & $5\,\mathrm{GHz}$, $T_1=1.5\,\mathrm{s}$ & $2\times10^{-48}$\,$^{b}$ & $10^{-38}$ & $3\times10^{-23}$ & $10^{-26}$\,$^{f}$ & -- & $2\times10^{66}$\,$^{f}$\\
Bulk heating, Si~\cite{AdlerVinante2018} & $\dot E/M\lesssim10^{-11}\,\mathrm{W/kg}$ & $5\times10^{-49}$\,$^{g}$ & $2\times10^{-56}$ & \multicolumn{2}{c}{$2\times10^{-44}$\,$^{h}$} & -- & $10^{49}$\\
\end{tabular}
\end{ruledtabular}
\begin{flushleft}
$^{a}$Correlation length between the test-mass size and the distance to the reference. $^{b}$Correlation length $\ell_c\sim\lambda/2$, with $m_A\sim\varrho(\lambda/2)^3$, and for the nanobeam the mass of a beam segment of length $\lambda/2$. Collective fluctuations do not couple to internal modes. $^{c}$Breuer position diffusion of a free mass at rest, $\tau_B\langle\hat p^2\rangle/M^2$, negligible. $^{d}$Velocity noise $\tau_Bk_BT_{\rm kin}/M$ from the atomic motion, with $T_{\rm kin}\simeq300$~K. $^{e}$Internal temperature $T_{\rm kin}\simeq1000$~K measured for a comparable particle~\cite{Hebestreit2018}, with bounds scaling as $T_{\rm kin}^{-1}$ and $T_{\rm kin}^{-3}$, respectively. For collective fluctuations, the composite value of $\tau_I$ is $3\times10^{75}$. $^{f}$$T_{\rm kin}$ set by zero-point motion, about $180$~K, $390$~K and $240$~K for quartz, sapphire and silicon. $^{g}$Gaussian correlations with $\ell_c=1\,\mu\mathrm{m}$, the bound scaling as $\ell_c^2$. $^{h}$Both independent and collective fluctuations. $^{i}$Measured, from the decay of phonon Fock and superposition states, treating the mode as an elementary oscillator of effective mass $16.2\,\mu\mathrm{g}$~\cite{donadi_quantum_2025}.
\end{flushleft}
\end{table*}

\subsection{Bulk heating of matter}
\label{sec:bulkHeating}

The bounds above concern the single mode that is read out. The same dissipators, however, act on every normal mode of a solid, and summing the energy injection over all modes gives the heating power of a body of mass $M$, which can be compared with calorimetric measurements. For a harmonic lattice with potential energy $U$, we obtain (see Eq.~\eqref{eq:generalCompositeDissipator})
\begin{subequations}
\label{eq:bulkHeatingGeneral}
\begin{align}
\dot E_A&=\frac{\tau_Ac^2}{2}\sum_{a,b}C_{ab}\,\partial_a\partial_bU,\\
\dot E_B&=\frac{\tau_B}{4m^2}\sum_{a,b}C_{ab}\,\partial_a\partial_bU\,\langle\{\hat{\mathbf p}_a,\hat{\mathbf p}_b\}\rangle ,
\end{align}
\end{subequations}
with $\partial_a$ the gradient with respect to the position of atom $a$ and summation over Cartesian components implied. Since these expressions involve the full atomic momenta, they include the internal-motion contributions of Appendix~\ref{sec:compositeMain}.

The Arzano contribution is independent of the state and of the temperature. For independent fluctuations, $\sum_a\partial_a^2U=m\sum_\kappa\omega_\kappa^2$, where $\kappa$ labels the $3N$ normal modes of frequency $\omega_\kappa$, so that for a Debye spectrum with Debye frequency $\omega_D$, i.e., $\langle\omega^2\rangle=3\omega_D^2/5$,
\begin{equation}
\frac{\dot E_A}{M}=\frac{3}{2}c^2\tau_A\langle\omega^2\rangle .
\label{eq:bulkArzanoIndep}
\end{equation}
For correlations of range $\ell_c$ much larger than the lattice constant, an acoustic mode of wave vector $\mathbf q$ contributes $m_A\omega^2c^2\tau_A/2$ with $m_A\simeq\varrho\tilde C(\mathbf q)$, where $\varrho$ is the mass density and $\tilde C$ the Fourier transform of $C(\mathbf r)$, so that only wavelengths $\gtrsim\ell_c$ contribute, and
\begin{equation}
\frac{\dot E_A}{M}=\frac{c^2\tau_A}{2}\,\big[-\nabla^2C\big](0)\,\big(v_L^2+2v_T^2\big)=\frac{3c^2\tau_A}{2\ell_c^2}\big(v_L^2+2v_T^2\big),
\label{eq:bulkArzanoCorr}
\end{equation}
with $v_L$ and $v_T$ the longitudinal and transverse sound velocities, the last equality holding for Gaussian correlations. For collective fluctuations, $\dot E_A=0$. For the Breuer model with collective fluctuations, each mode is heated as in Eq.~\eqref{eq:BreuerHeatingEnergy}, so that $\dot E_B=\tau_B\sum_\kappa\omega_\kappa^2\langle\hat K_\kappa\rangle$, with $\hat K_\kappa$ the kinetic energy of mode $\kappa$. 
Assuming the phonon modes are in thermal equilibrium, \(\langle\hat K_\kappa\rangle\geq\hbar\omega_\kappa/4\), so zero-point motion alone gives the following lower bound at any temperature
\begin{equation}
\frac{\dot E_B}{M}\geq\frac{3\tau_B\hbar\langle\omega^3\rangle}{4m}=\frac{3\tau_B\hbar\omega_D^3}{8m}.
\label{eq:bulkBreuer}
\end{equation}
Since $\partial_a\partial_bU$ couples only neighboring atoms, $\dot E_B$ depends only on the correlations between neighbors, and for independent fluctuations it changes only by a factor of order unity.

After accounting for radioactivity and cosmic-ray muons, residual heating in solids at low temperatures is $\dot E/M\lesssim10^{-11}\,\mathrm{W/kg}$ \cite{AdlerVinante2018}. For silicon, with $\omega_D\simeq8.4\times10^{13}\,\mathrm{s^{-1}}$, $v_L\simeq8.4\,\mathrm{km/s}$ and $v_T\simeq5.8\,\mathrm{km/s}$, Eqs.~\eqref{eq:bulkArzanoIndep}--\eqref{eq:bulkBreuer} give
\begin{align}
\tau_A&\lesssim2\times10^{-56}\,\mathrm{s},\qquad \tau_A\lesssim5\times10^{-49}\,\mathrm{s}\left(\frac{\ell_c}{1\,\mu\mathrm{m}}\right)^{2},\nonumber\\
\tau_B&\lesssim2\times10^{-44}\,\mathrm{s}\simeq0.4\,t_P,
\label{eq:bulkBounds}
\end{align}
for independent fluctuations, Gaussian correlations of range $\ell_c$, and both independent/collective fluctuations, respectively. 
The first bound replaces the second for $\ell_c$ of the order of the lattice constant. For the Petruzziello--Illuminati model with independent fluctuations, treating each atom as an oscillator of frequency $3\omega_D/4$ in its ground state, with the rate of Eq.~\eqref{eq:IlluminatiInitialHeatingGround}, gives $\tau_I\lesssim10^{49}\,\mathrm{s}$, i.e., $\sigma_\beta\lesssim10^{46}$ for $\tau_c=t_P$, the tightest bound on this model.

These bounds assume that the constituents are atoms, that the noise is white up to $\omega_D$, and that the master equations hold in their Lindblad form, which corresponds to an infinite-temperature noise. Resolving nucleons or electrons would increase the rates further, whereas dissipative extensions, as considered for collapse models~\cite{AdlerVinante2018}, would cap the heating at an effective noise temperature.

\subsection{Overview}

Table~\ref{tab:bounds} shows that existing mechanical platforms can already probe Planck-scale phenomenology for the Arzano and Breuer models, under assumptions on the correlation structure that the table makes explicit. 
For the Arzano model, free test masses reach $\tau_A\ll t_P$ if the fluctuations are correlated over distances larger than the test masses but shorter than their separation, acoustic resonators would do so for micrometer-scale correlations, and bulk heating gives the tightest bound for shorter correlation lengths, down to $\tau_A\lesssim2\times10^{-56}\,\mathrm{s}$ for independent fluctuations. Different platforms thus constrain complementary ranges of $\ell_c$, turning the main ambiguity of the composite-body extension into a measurable property. For the Breuer model, fluctuation--dissipation tests of ultracoherent resonators and residual-occupation measurements of gigahertz modes could improve on ground-state heating of levitated particles by eight to fourteen orders of magnitude, independent fluctuations are constrained even more tightly through the zero-point motion of the atoms, and bulk heating reaches $\tau_B\lesssim0.4\,t_P$ for both independent and collective fluctuations. For the Petruzziello--Illuminati model, the bounds remain far from the Planck scale in both interpretations. Thermal fluctuation--dissipation tests of elementary modes and bulk heating improve on ground-state heating by more than thirty orders of magnitude, whereas coherent tests constrain the mean value of the GUP parameter far more tightly.

\section{Conclusions}

We have investigated how three phenomenological models of gravity-related decoherence, built on deformed relativistic symmetries, stochastic metric fluctuations and a stochastic generalized uncertainty principle, manifest themselves in observables that do not require the preparation of massive spatial superposition states. Their master equations share a double-commutator structure, but their collapse operators involve different powers of momentum, and this difference results in remarkably different predictions.

For a free particle, the three dissipators commute with the Hamiltonian, so that the momentum distribution and the energy are conserved. Their effects remain visible in the decay of momentum-space coherences and in an additional diffusion of the position, with rates whose dependence on the mean momentum differs among the models. Preparing states of different momentum can therefore discriminate between them, beyond a single overall decoherence rate.

For a harmonic oscillator, the collapse operators no longer commute with the Hamiltonian, and the models result in anomalous heating effects. The Arzano model heats the oscillator at a constant rate, the Breuer model at a rate proportional to the kinetic energy, leading to exponential growth below a damping threshold, and the Petruzziello--Illuminati model at a rate growing cubically with excitation, with the secular approximation predicting finite-time divergence to arbitrarily high occupation. A stochastic unraveling traces this hierarchy to a velocity noise whose momentum dependence is constant, linear and cubic, respectively. In the experimentally relevant regime of damping stronger than the gravity-related decoherence, all three models admit a stationary (or quasistationary) state, whose leading-order position spectra share the same excess lineshape. This lineshape distinguishes them from conventional force noise, but not from each other, which requires combining spectra with heating and decoherence measurements.

Applying these results to real oscillators requires extending the single-particle master equations to composite bodies, which is not unique. The resulting rates depend on how the fluctuations are correlated between atoms and, except for the Arzano model, on the thermal and zero-point motion of the atoms, which can dominate over the motion of the mode itself. Within this framework, existing displacement-noise and calorimetric measurements already constrain the Arzano model well below the Planck time, and the Breuer model at the Planck time, while mechanical modes and levitated particles probe complementary ranges of correlation length.

These strategies rely only on standard tools of mechanical sensing, namely ringdown, thermometry, and continuous displacement measurements. In the case of an anomalous signal, the distinct dependences on excitation, frequency, mass, correlation length and spectral shape would allow the three mechanisms to be told apart from each other and from technical noise. Mechanical oscillators thus provide a setting, complementary to matter-wave interferometry and tests of gravitationally induced entanglement, in which competing low-energy signatures of quantum gravity can be constrained and compared.

\vspace{5mm}
\textbf{Acknowledgments. --} SD acknowledges support from University of Trieste and INFN. IPL was partially supported by the National Council for Scientific and Technological Development - CNPq, grant 312547/2023-4. MF was supported by the Swiss National Science Foundation Ambizione Grant No. 208886, and by The Branco Weiss Fellowship -- Society in Science, administered by the ETH Z\"{u}rich. 
The authors would like to acknowledge networking support by the COST Action CA23130 ``BridgeQG: Bridging high and low energies in search of quantum gravity'' and the COST Action CA23115 ``RQI: Relativistic Quantum Information''.

\bibliography{ref-fluc}

\appendix
\onecolumngrid

\clearpage
\newpage

\section{Connections to other models}

\subsection{Connection to Blencowe's model}
\label{suppsec:blencowe}

For a free particle, $\hat H_0=\hat K$, the Breuer dissipator is diagonal in the energy eigenbasis, $\hat H_0\ket{E}=E\ket{E}$. Equation~\eqref{eq:BreuerModel} then gives
\begin{equation}
\frac{d}{dt}\rho_{EE'}(t)=\left[-\frac{i}{\hbar}(E-E')-\frac{\tau_B}{2\hbar^2}(E-E')^2\right]\rho_{EE'}(t),
\label{eq:gamma_screenshot}
\end{equation}
so that the populations $\rho_{EE}$ are unchanged, whereas coherences between states of different energies decay exponentially with rate $\Gamma_{EE'}=\tau_B(E-E')^2/(2\hbar^2)$. The Breuer term therefore describes pure dephasing in the kinetic-energy basis for a free particle.

A closely related energy-dephasing law was obtained by Blencowe~\cite{Blencowe:2012mp} from an effective-field-theory treatment of weak, quantized metric perturbations. In that approach the gravitational field is treated as an environment in a thermal state at temperature $T$ and is traced out from the matter dynamics. In the Born--Markov and high-temperature limits, Blencowe finds for a stationary superposition of matter states with energies $E$ and $E'$ the decoherence rate [Eq.~(19) of Ref.~\cite{Blencowe:2012mp}]
\begin{equation}
\Gamma_{EE'}^{(\mathrm{Bl})}=\frac{k_B T}{\hbar}\left(\frac{E-E'}{E_P}\right)^2,
\label{eq:gamma_blencowe}
\end{equation}
where $E_P=\sqrt{\hbar c^5/G}$ is the Planck energy and $T$ characterizes the thermal graviton background, rather than the temperature of the matter system. The quadratic dependence on $E-E'$ has a simple physical interpretation: fluctuations of the gravitational environment generate fluctuations of the relative phase accumulated by the two branches, and averaging over these fluctuations suppresses their coherence. In particular, within the stationary limit considered by Blencowe, two spatially distinct states with the same energy do not decohere through this mechanism.

Comparing Eqs.~\eqref{eq:gamma_screenshot} and \eqref{eq:gamma_blencowe} at the level of their energy-basis dephasing rates gives
\begin{equation}
\tau_B=2\hbar\,\frac{k_B T}{E_P^2}.
\label{eq:tauB_map}
\end{equation}
This identification should be understood as a correspondence between the two decoherence rates, rather than as a microscopic equivalence of the models. In Blencowe's derivation the relevant energy of the stationary matter configurations is dominated by their rest energy, whereas in the nonrelativistic free-particle Breuer model considered here $E$ denotes the kinetic energy. Nevertheless, both mechanisms share the characteristic $\Gamma\propto(E-E')^2$ energy-dephasing structure.

\section{Exact solution of the Arzano model}
\label{Arzano_exact}

In this section we solve exactly the Arzano master equation for a one-dimensional harmonic oscillator in the presence of zero-temperature amplitude damping. We first solve the closed equations for the second moments, which determine the heating dynamics and the stationary spectrum. We then obtain the full solution for an arbitrary initial state in terms of its Weyl characteristic function, and use it to determine the evolution of Fock-state populations and of the coherence $\langle0|\hat\rho(t)|1\rangle$ for the initial state $(|0\rangle+|1\rangle)/\sqrt{2}$.

\subsection{Exact second-moment dynamics and heating}

Let us define the variables $E=K+U$ and $W=K-U$. In the variables $(E,W,S)$, Eqs.~\eqref{eq:ArzanoMoments} read
\begin{align}
\dot E&=\frac{\hbar\omega}{2}\Gamma^{\rm HO}_A+\gamma\left(\frac{\hbar\omega}{2}-E\right),\label{eq:S4_A_E}\\
\dot W&=-\omega^2 S-\frac{\hbar\omega}{2}\Gamma^{\rm HO}_A-\gamma W,\label{eq:S4_A_D}\\
\dot S&=4W-\gamma S.\label{eq:S4_A_S}
\end{align}
The total energy decouples. Its solution for an arbitrary initial state is
\begin{equation}
E(t)=E_\infty+\left(E_0-E_\infty\right)e^{-\gamma t},\qquad E_\infty=\frac{\hbar\omega}{2}\left(1+\frac{\Gamma^{\rm HO}_A}{\gamma}\right),
\end{equation}
which gives Eqs.~\eqref{eq:ArzanoNumberSolution} and \eqref{eq:ArzanoHeatingStationary} of the main text. For $\gamma\to0$ it reduces to $E(t)=E_0+\hbar\omega\Gamma^{\rm HO}_A t/2$, Eq.~\eqref{eq:ArzanoLinearHeating}.

The pair $(W,S)$ obeys a linear system with drift matrix $\left(\begin{smallmatrix}-\gamma&-\omega^2\\4&-\gamma\end{smallmatrix}\right)$, whose eigenvalues are $-\gamma\pm2i\omega$, and with a constant source. Its fixed point is
\begin{equation}
S_{\rm ss}=-\frac{2\hbar\omega\Gamma^{\rm HO}_A}{\gamma^2+4\omega^2},\qquad W_{\rm ss}=\frac{\gamma}{4}S_{\rm ss}=-\frac{\gamma\hbar\omega\Gamma^{\rm HO}_A}{2(\gamma^2+4\omega^2)}.
\end{equation}
Writing $w_0=W_0-W_{\rm ss}$ and $s_0=S_0-S_{\rm ss}$, the exact solution is
\begin{align}
W(t)&=W_{\rm ss}+e^{-\gamma t}\left[w_0\cos(2\omega t)-\frac{\omega}{2}s_0\sin(2\omega t)\right],\\
S(t)&=S_{\rm ss}+e^{-\gamma t}\left[s_0\cos(2\omega t)+\frac{2}{\omega}w_0\sin(2\omega t)\right].
\end{align}
Eliminating $W$ gives Eq.~\eqref{eq:ArzanoMixedMoment} of the main text. The kinetic and potential energies follow from $K=(E+W)/2$ and $U=(E-W)/2$. In particular,
\begin{equation}
\langle\hat x^2\rangle_{\rm ss}=\frac{E_\infty-W_{\rm ss}}{m\omega^2}=\frac{\hbar}{2m\omega}+\frac{\hbar\Gamma^{\rm HO}_A}{m\omega}\,\frac{\gamma^2+2\omega^2}{\gamma(\gamma^2+4\omega^2)},
\end{equation}
which, together with $S_{\rm ss}$, gives Eq.~\eqref{eq:ArzanoSpectrumMoments} and hence the spectrum~\eqref{eq:ArzanoSpectrum}.

\subsection{Master equation}

The dynamics is governed by
\begin{equation}
\partial_t\hat\rho=-\frac{i}{\hbar}[\hat H_0,\hat\rho]-\tau_A\frac{c^2}{2\hbar^2}[\hat p,[\hat p,\hat\rho]]+\gamma\mathcal D[\hat a]\hat\rho,\qquad \hat H_0=\hbar\omega\left(\hat N+\frac12\right),\quad \hat N=\hat a^\dagger\hat a,
\label{eq:initial-master}
\end{equation}
where $\mathcal D[\hat a]\hat\rho=\hat a\hat\rho\hat a^\dagger-\frac12\{\hat a^\dagger\hat a,\hat\rho\}$. Introducing the dimensionless quadratures
\begin{equation}
\hat Q=\frac{\hat a+\hat a^\dagger}{\sqrt2},\qquad \hat P=\frac{\hat a-\hat a^\dagger}{i\sqrt2},\qquad [\hat Q,\hat P]=i,
\end{equation}
one has $\hat p=\sqrt{m\hbar\omega}\,\hat P$. Since $\mathcal D[\hat P]\hat\rho=-[\hat P,[\hat P,\hat\rho]]/2$, Eq.~\eqref{eq:initial-master} becomes
\begin{equation}
\dot{\hat\rho}=-i\omega[\hat N,\hat\rho]+\gamma\mathcal D[\hat a]\hat\rho+\Gamma_A^{\rm HO}\mathcal D[\hat P]\hat\rho,
\label{eq:compact-master}
\end{equation}
where $\Gamma_A^{\rm HO}=m\omega c^2\tau_A/\hbar$ is the characteristic Arzano rate introduced in Eq.~\eqref{eq:GammaAHO}.

\subsection{Characteristic-function equation}

We introduce the Weyl characteristic function
\begin{equation}
\chi_t(k_q,k_p)=\operatorname{Tr}\left[\hat\rho(t)e^{i(k_q\hat Q+k_p\hat P)}\right].
\label{eq:chi-definition}
\end{equation}
Applying Eq.~\eqref{eq:compact-master} to the Weyl operator gives
\begin{equation}
\partial_t\chi_t=\left(-\frac{\gamma}{2}k_q-\omega k_p\right)\partial_{k_q}\chi_t+\left(\omega k_q-\frac{\gamma}{2}k_p\right)\partial_{k_p}\chi_t-\left[\frac{\gamma}{4}(k_q^2+k_p^2)+\frac{\Gamma_A^{\rm HO}}{2}k_q^2\right]\chi_t.
\label{eq:chi-pde}
\end{equation}
The amplitude-damping channel therefore produces isotropic diffusion in the characteristic variables, whereas the Arzano term contributes only along $k_q$, as expected for a dissipator generated by $\hat P$.

Defining $\bm k=(k_q,k_p)^T$ and $\nabla_{\bm k}=(\partial_{k_q},\partial_{k_p})^T$, Eq.~\eqref{eq:chi-pde} can be written as
\begin{equation}
\partial_t\chi_t(\bm k)=(B\bm k)\cdot\nabla_{\bm k}\chi_t(\bm k)-\frac12\bm k^T\mathsf D\bm k\,\chi_t(\bm k),
\label{eq:chi-matrix}
\end{equation}
with
\begin{equation}
B=\begin{pmatrix}-\gamma/2&-\omega\\ \omega&-\gamma/2\end{pmatrix},\qquad \mathsf D=\begin{pmatrix}\gamma/2+\Gamma_A^{\rm HO}&0\\0&\gamma/2\end{pmatrix}.
\end{equation}

\subsection{General solution}

The drift propagator is
\begin{equation}
\mathsf R(t)\equiv e^{Bt}=e^{-\gamma t/2}\begin{pmatrix}\cos(\omega t)&-\sin(\omega t)\\ \sin(\omega t)&\cos(\omega t)\end{pmatrix}.
\label{eq:U-definition}
\end{equation}
To solve Eq.~\eqref{eq:chi-matrix} by the method of characteristics, we introduce a constant vector $\bm\kappa$ and the trajectory $\bm k(t)=\mathsf R(-t)\bm\kappa$, which satisfies $\dot{\bm k}=-B\bm k$. Along this trajectory, $\widetilde\chi_t(\bm\kappa)\equiv\chi_t[\mathsf R(-t)\bm\kappa]$ obeys
\begin{equation}
\partial_t\widetilde\chi_t(\bm\kappa)=-\frac12\bm\kappa^T\mathsf R(-t)^T\mathsf D\mathsf R(-t)\bm\kappa\,\widetilde\chi_t(\bm\kappa).
\label{eq:chi-tilde-ode}
\end{equation}
Integration gives
\begin{equation}
\widetilde\chi_t(\bm\kappa)=\chi_0(\bm\kappa)\exp\left[-\frac12\int_0^t\mathrm ds\,\bm\kappa^T\mathsf R(-s)^T\mathsf D\mathsf R(-s)\bm\kappa\right].
\label{eq:chi-tilde-solution}
\end{equation}
Setting $\bm\kappa=\mathsf R(t)\bm k$ and using $\mathsf R(-s)\mathsf R(t)=\mathsf R(t-s)$ then yields the exact solution
\begin{equation}
\chi_t(\bm k)=\chi_0[\mathsf R(t)\bm k]\exp\left[-\frac12\bm k^TY(t)\bm k\right],\qquad Y(t)=\int_0^t\mathrm du\,\mathsf R(u)^T\mathsf D\mathsf R(u),
\label{eq:generic-solution}
\end{equation}
valid for an arbitrary initial characteristic function $\chi_0$.

Separating the diffusion matrix as $\mathsf D=(\gamma/2)\id_2+\Gamma_A^{\rm HO}\operatorname{diag}(1,0)$ and using $\mathsf R(u)^T\mathsf R(u)=e^{-\gamma u}\id_2$, one finds
\begin{equation}
Y(t)=\frac{1-e^{-\gamma t}}{2}\id_2+\Sigma_A(t),
\label{eq:Y-split}
\end{equation}
where the Arzano-induced contribution is
\begin{equation}
\Sigma_A(t)=\Gamma_A^{\rm HO}\int_0^t\mathrm du\,e^{-\gamma u}\begin{pmatrix}\cos^2(\omega u)&-\cos(\omega u)\sin(\omega u)\\-\cos(\omega u)\sin(\omega u)&\sin^2(\omega u)\end{pmatrix}.
\label{eq:S-integral}
\end{equation}
Consequently,
\begin{equation}
\chi_t(\bm k)=\chi_0[\mathsf R(t)\bm k]\exp\left[-\frac{1-e^{-\gamma t}}{4}\bm k^T\bm k-\frac12\bm k^T\Sigma_A(t)\bm k\right].
\label{eq:generic-solution-explicit}
\end{equation}

Writing $\Sigma_A(t)=\left(\begin{smallmatrix}\sigma_{qq}(t)&\sigma_{qp}(t)\\\sigma_{qp}(t)&\sigma_{pp}(t)\end{smallmatrix}\right)$, its components are
\begin{equation}
\sigma_{qq}=\frac{\Gamma_A^{\rm HO}}{2}(I_0+I_c),\qquad \sigma_{pp}=\frac{\Gamma_A^{\rm HO}}{2}(I_0-I_c),\qquad \sigma_{qp}=-\frac{\Gamma_A^{\rm HO}}{2}I_s,
\label{eq:S-components}
\end{equation}
where
\begin{equation}
I_0(t)=\frac{1-e^{-\gamma t}}{\gamma},\qquad I_c(t)=\frac{\gamma+e^{-\gamma t}[-\gamma\cos(2\omega t)+2\omega\sin(2\omega t)]}{\gamma^2+4\omega^2},\qquad I_s(t)=\frac{2\omega-e^{-\gamma t}[\gamma\sin(2\omega t)+2\omega\cos(2\omega t)]}{\gamma^2+4\omega^2}.
\end{equation}
These expressions have regular limits for $\gamma\to0$ and completely determine the Gaussian channel generated by Eq.~\eqref{eq:compact-master}.

\subsection{Initial state $\ket{\psi_{0+1}}=(\ket{0}+\ket{1})/\sqrt{2}$}

We first consider the initial state used to probe the $|0\rangle$--$|1\rangle$ coherence in the main text, $\hat\rho_{0+1}(0)=\ket{\psi_{0+1}}\bra{\psi_{0+1}}$. Defining $k^2\equiv k_q^2+k_p^2$, its characteristic function is
\begin{equation}
\chi_0(\bm k)=e^{-k^2/4}\left(1-\frac{k^2}{4}+\frac{ik_q}{\sqrt2}\right).
\label{eq:initial-chi}
\end{equation}
Since $\lVert\mathsf R(t)\bm k\rVert^2=e^{-\gamma t}k^2$ and $[\mathsf R(t)\bm k]_q=e^{-\gamma t/2}[k_q\cos(\omega t)-k_p\sin(\omega t)]$, the general solution~\eqref{eq:generic-solution-explicit} gives
\begin{equation}
\chi_t(\bm k)=\exp\left[-\frac{k^2}{4}-\frac12\bm k^T\Sigma_A(t)\bm k\right]\left[1-\frac{e^{-\gamma t}}{4}k^2+\frac{ie^{-\gamma t/2}}{\sqrt2}\bigl(k_q\cos\omega t-k_p\sin\omega t\bigr)\right].
\label{eq:specific-chi}
\end{equation}

The density operator is recovered from $\hat\rho(t)=(2\pi)^{-1}\int dk_q\,dk_p\,\chi_t(\bm k)e^{-i(k_q\hat Q+k_p\hat P)}$. Using $\langle0|e^{-i(k_q\hat Q+k_p\hat P)}|1\rangle=-(k_p+ik_q)e^{-k^2/4}/\sqrt2$ and evaluating the resulting Gaussian integral, one obtains
\begin{equation}
\langle0|\hat\rho_{0+1}(t)|1\rangle=\frac{e^{-\gamma t/2}}{2\Delta_A^{3/2}(t)}\left\{[1+\sigma_{pp}(t)]\cos(\omega t)+\sigma_{qp}(t)\sin(\omega t)+i\left[\sigma_{qp}(t)\cos(\omega t)+[1+\sigma_{qq}(t)]\sin(\omega t)\right]\right\},
\label{eq:coherence}
\end{equation}
where
\begin{equation}
\Delta_A(t)\equiv\det[\id_2+\Sigma_A(t)]=[1+\sigma_{qq}(t)][1+\sigma_{pp}(t)]-\sigma_{qp}^2(t).
\label{eq:DeltaA}
\end{equation}
Equation~\eqref{eq:coherence} is exact in $\Gamma_A^{\rm HO}$, $\gamma t$, and $\omega t$. For $\Gamma_A^{\rm HO}=0$ it reduces to $e^{i\omega t}e^{-\gamma t/2}/2$, as expected for amplitude damping alone.

For comparison with the perturbative expressions used in the main text, we expand to first order in $\Gamma_A^{\rm HO}$. Using $\sigma_{qq}+\sigma_{pp}=\Gamma_A^{\rm HO}(1-e^{-\gamma t})/\gamma$, we find
\begin{equation}
\langle0|\hat\rho_{0+1}(t)|1\rangle=\frac{e^{-\gamma t/2}}{2}\left[e^{i\omega t}\left(1-\frac{\Gamma_A^{\rm HO}}{\gamma}[1-e^{-\gamma t}]\right)-\frac{\Gamma_A^{\rm HO}}{2}\frac{e^{-i\omega t}-e^{-\gamma t}e^{i\omega t}}{\gamma-2i\omega}\right]+O[(\Gamma_A^{\rm HO})^2].
\label{eq:coherence-first-order}
\end{equation}
This is an expansion only in the strength of the Arzano term and does not assume $\gamma t\ll1$ or $\omega t\ll1$. The second term in the square brackets is of order $\Gamma_A^{\rm HO}/\omega$ and does not grow in time, so that for $\gamma t\ll1\ll\omega t$ the secular part reduces to Eq.~\eqref{eq:ArzanoHOcoherence}.

As a useful check, in the undamped limit $\gamma\to0$,
\begin{equation}
\langle0|\hat\rho_{0+1}(t)|1\rangle=\frac12e^{i\omega t}(1-\Gamma_A^{\rm HO}t)-\frac{\Gamma_A^{\rm HO}}{4\omega}\sin(\omega t)+O[(\Gamma_A^{\rm HO})^2].
\label{eq:coherence-undamped}
\end{equation}
The effective coherence-loss rate therefore depends on the time window. For $\omega^{-1}\ll t\ll\gamma^{-1}$ the oscillating term averages out and the coherence decays at the secular rate $\Gamma_A^{\rm HO}$ of Eq.~\eqref{eq:ArzanoHOcoherence}, the mean of the population-loss rates $\Gamma_A^{\rm HO}/2$ and $3\Gamma_A^{\rm HO}/2$ of $\ket0$ and $\ket1$. For $t\ll\gamma^{-1},\omega^{-1}$, instead, the non-secular term contributes at the same order and
\begin{equation}
\langle0|\hat\rho_{0+1}(t)|1\rangle\simeq\frac12e^{-\gamma t/2}e^{i\omega t}\left(1-\frac32\Gamma_A^{\rm HO}t\right).
\label{eq:coherence-very-short-time}
\end{equation}

\subsection{Initial state $\ket0$}

For an oscillator initially in its ground state, $\hat\rho_0(0)=\ket0\bra0$, the characteristic function is $\chi_0^{(0)}(\bm k)=e^{-k^2/4}$. Equation~\eqref{eq:generic-solution-explicit} therefore gives
\begin{equation}
\chi_t^{(0)}(\bm k)=\exp\left[-\frac{k^2}{4}-\frac12\bm k^T\Sigma_A(t)\bm k\right].
\label{eq:chi-0}
\end{equation}
The absence of any additional $\gamma$-dependent factor reflects the fact that the vacuum is a fixed state of zero-temperature amplitude damping. The damping rate nevertheless enters through $\Sigma_A(t)$, because it competes with the Arzano-induced diffusion. Using the vacuum matrix element of the Weyl operator and performing the Gaussian integral gives the exact survival probability
\begin{equation}
\langle0|\hat\rho_0(t)|0\rangle=\frac{1}{\sqrt{\Delta_A(t)}}.
\label{eq:rho00-exact}
\end{equation}
Since $\Sigma_A(t)=O(\Gamma_A^{\rm HO})$, expansion to first order yields
\begin{equation}
\langle0|\hat\rho_0(t)|0\rangle=1-\frac{\Gamma_A^{\rm HO}}{2\gamma}\left(1-e^{-\gamma t}\right)+O[(\Gamma_A^{\rm HO})^2],
\label{eq:rho00-first-order}
\end{equation}
whose short-time limit is the first of Eqs.~\eqref{eq:ArzanoHOpopulations}. In the absence of damping this becomes $\langle0|\hat\rho_0(t)|0\rangle=1-\Gamma_A^{\rm HO}t/2+O[(\Gamma_A^{\rm HO})^2]$. Thus, unlike ordinary zero-temperature amplitude damping, the Arzano dissipator drives an initially prepared ground state out of the vacuum, with initial depletion rate $\Gamma_A^{\rm HO}/2$.

\subsection{Initial state $\ket1$}

For $\hat\rho_1(0)=\ket1\bra1$, the initial characteristic function is
\begin{equation}
\chi_0^{(1)}(\bm k)=e^{-k^2/4}\left(1-\frac{k^2}{2}\right),
\label{eq:initial-chi-1}
\end{equation}
and its evolution follows directly from Eq.~\eqref{eq:generic-solution-explicit},
\begin{equation}
\chi_t^{(1)}(\bm k)=\exp\left[-\frac{k^2}{4}-\frac12\bm k^T\Sigma_A(t)\bm k\right]\left(1-\frac{e^{-\gamma t}}{2}k^2\right).
\label{eq:chi-1}
\end{equation}
Using $\langle1|e^{-i(k_q\hat Q+k_p\hat P)}|1\rangle=e^{-k^2/4}(1-k^2/2)$, the survival probability is again obtained from Gaussian moments. Defining $u_A(t)\equiv2+\sigma_{qq}(t)+\sigma_{pp}(t)$, the exact result is
\begin{equation}
\langle1|\hat\rho_1(t)|1\rangle=\frac{1}{\sqrt{\Delta_A(t)}}\left[1-\frac{[1+e^{-\gamma t}]u_A(t)}{2\Delta_A(t)}-\frac{e^{-\gamma t}}{\Delta_A(t)}+\frac{3e^{-\gamma t}u_A^2(t)}{4\Delta_A^2(t)}\right].
\label{eq:rho11-exact}
\end{equation}
For $\Gamma_A^{\rm HO}=0$, one has $\Sigma_A=0$, $\Delta_A=1$, and $u_A=2$, recovering $\langle1|\hat\rho_1(t)|1\rangle=e^{-\gamma t}$ as required. Expanding the exact result to first order in the Arzano rate gives
\begin{equation}
\langle1|\hat\rho_1(t)|1\rangle=e^{-\gamma t}+\frac{\Gamma_A^{\rm HO}}{2\gamma}\left(1-e^{-\gamma t}\right)\left(1-4e^{-\gamma t}\right)+O[(\Gamma_A^{\rm HO})^2],
\label{eq:rho11-first-order}
\end{equation}
whose leading short-time form is the second of Eqs.~\eqref{eq:ArzanoHOpopulations}. In the undamped limit this reduces to $\langle1|\hat\rho_1(t)|1\rangle=1-\frac32\Gamma_A^{\rm HO}t+O[(\Gamma_A^{\rm HO})^2]$, while for $\gamma t\ll1$,
\begin{equation}
\langle1|\hat\rho_1(t)|1\rangle\simeq1-\left(\gamma+\frac32\Gamma_A^{\rm HO}\right)t.
\label{eq:rho11-short-time}
\end{equation}
The two contributions describe different processes: amplitude damping transfers population from $\ket1$ to $\ket0$, whereas the Hermitian Arzano collapse operator $\hat P\propto\hat a-\hat a^\dagger$ induces transitions from $\ket1$ toward both lower and higher Fock levels, at rates $\Gamma_A^{\rm HO}/2$ and $\Gamma_A^{\rm HO}$, respectively. At long times, the term $\Gamma_A^{\rm HO}/2\gamma$ in Eq.~\eqref{eq:rho11-first-order} reflects the population of $\ket1$ maintained by the balance between Arzano-induced excitation from the ground state and amplitude damping.

\section{Exact solution of the Breuer model}
\label{app:BreuerExact}

The equations of motion for the Breuer model are Eqs.~\eqref{eq:BreuerMoments}. The essential difference with respect to the Arzano model is that the gravity-related contribution to $\dot U$ is not a constant source but is proportional to $K$. Adding the equations for $K$ and $U$ gives Eq.~\eqref{eq:BreuerHeatingEnergy},
\begin{equation}
\dot E_t=\Gamma^{\rm HO}_B K+\gamma\left(\frac{\hbar\omega}{2}-E_t\right),
\end{equation}
so the energy equation does not close by itself. As the kinetic energy grows, the energy injection grows with it, which opens the possibility of an exponential instability.

\subsection{Equation for the kinetic energy}
It is convenient to introduce the operator $\mathcal{T}\equiv d/dt+\gamma$. With it, Eqs.~\eqref{eq:BreuerMoments} read
\begin{equation}
\mathcal T K=-\frac{\omega^2}{2}S+\frac{\gamma\hbar\omega}{4},\qquad \mathcal T U=\frac{\omega^2}{2}S+\Gamma^{\rm HO}_BK+\frac{\gamma\hbar\omega}{4},\qquad \mathcal T S=4(K-U).
\end{equation}
Since $\mathcal T$ acting on a constant $c$ gives $\gamma c$, applying $\mathcal T$ twice to the equation for $K$ gives
\begin{align}
\mathcal T^2K&=-2\omega^2(K-U)+\gamma^2\frac{\hbar\omega}{4},\\
\mathcal T^3K&=-2\omega^2(\mathcal TK-\mathcal TU)+\gamma^3\frac{\hbar\omega}{4}=2\omega^4S+2\Gamma^{\rm HO}_B\omega^2K+\gamma^3\frac{\hbar\omega}{4}.
\end{align}
Eliminating $S$ with $\omega^2S=-2\mathcal TK+\gamma\hbar\omega/2$, we obtain a closed third-order equation for the kinetic energy,
\begin{equation}
\mathcal T^3K+4\omega^2\,\mathcal TK-2\Gamma^{\rm HO}_B\omega^2K=\frac{\gamma\hbar\omega}{4}\left(\gamma^2+4\omega^2\right).
\label{eq:S4_B_Kode}
\end{equation}
Once $K(t)$ is known, the other two moments follow algebraically from the equations for $\dot K$ and $\dot S$:
\begin{equation}
S=\frac{2}{\omega^2}\left[\gamma\left(\frac{\hbar\omega}{4}-K\right)-\dot K\right],\qquad U=K-\frac{1}{4}\left(\dot S+\gamma S\right).
\label{eq:S4_B_SU_from_K}
\end{equation}

\subsection{Characteristic exponents}
The homogeneous solutions of Eq.~\eqref{eq:S4_B_Kode} are $e^{(z-\gamma)t}$, where $z$ is a root of
\begin{equation}
P(z)\equiv z^3+4\omega^2 z-2\Gamma^{\rm HO}_B\omega^2=0.
\label{eq:S4_B_P}
\end{equation}
This is the characteristic polynomial of the undamped matrix in Eq.~\eqref{eq:BreuerMatrix}, and it gives Eq.~\eqref{eq:rBreuer}. Damping therefore shifts every eigenvalue of the undamped problem by $-\gamma$ without changing the corresponding eigenvectors.

Since $P'(z)=3z^2+4\omega^2>0$ for real $z$ and $P(0)=-2\Gamma^{\rm HO}_B\omega^2<0$, Eq.~\eqref{eq:S4_B_P} has a unique real root, and this root, $r$, is positive. The coefficient of $z^2$ vanishes, so the three roots add up to zero and the other two are $-r/2\pm i\Omega$. Their product, $r(r^2/4+\Omega^2)=2\Gamma^{\rm HO}_B\omega^2=r^3+4\omega^2r$, fixes
\begin{equation}
\Omega=\frac{1}{2}\sqrt{3r^2+16\omega^2},
\end{equation}
which is Eq.~\eqref{eq:OmegaBreuer}. The three characteristic exponents of the damped system are
\begin{equation}
\lambda_1=r-\gamma,\qquad \lambda_{2,3}=\lambda_2\pm i\Omega,\qquad \lambda_2=-\gamma-\frac{r}{2}.
\label{eq:S4_B_exponents}
\end{equation}

\subsection{Stationary solution}
The constant term on the right-hand side of Eq.~\eqref{eq:S4_B_Kode} gives the particular solution $K_\infty=\gamma\hbar\omega(\gamma^2+4\omega^2)/[4P(\gamma)]$. Equations~\eqref{eq:S4_B_SU_from_K} then give $S_\infty$ and $U_\infty$. Introducing
\begin{equation}
\mathcal D_B\equiv P(\gamma)=\gamma^3+4\gamma\omega^2-2\Gamma^{\rm HO}_B\omega^2,
\end{equation}
one finds
\begin{subequations}\label{eq:S4_B_stationary}
\begin{align}
K_\infty&=\frac{\gamma\hbar\omega(\gamma^2+4\omega^2)}{4\mathcal D_B},\\
U_\infty&=\frac{\gamma\hbar\omega(\gamma^2+\gamma\Gamma^{\rm HO}_B+4\omega^2)}{4\mathcal D_B},\\
S_\infty&=-\frac{\gamma\hbar\omega\,\Gamma^{\rm HO}_B}{\mathcal D_B}.
\end{align}
\end{subequations}
These are Eqs.~\eqref{eq:BreuerStationary} of the main text. For $\Gamma^{\rm HO}_B\to0$ they reduce to $K_\infty=U_\infty=\hbar\omega/4$ and $S_\infty=0$, as expected for zero-temperature amplitude damping. Since $P$ is monotonically increasing and $P(r)=0$, one has $\mathcal D_B>0$ if and only if $\gamma>r$. The stationary solution is therefore physical and is approached at long times only in the stable regime $\gamma>r$. For $\gamma<r$ it is an unstable fixed point with negative $K_\infty$, and for $\gamma=r$ it does not exist.

\subsection{General solution}
For $\gamma\neq r$, the complete solution contains three independent integration constants $k_1$, $k_2$, $k_3$. The kinetic energy takes the form
\begin{equation}
K(t)=K_\infty+k_1e^{\lambda_1t}+e^{\lambda_2t}\left[k_2\cos(\Omega t)+k_3\sin(\Omega t)\right].
\label{eq:S4_B_Ksol}
\end{equation}
We insert Eq.~\eqref{eq:S4_B_Ksol} into Eqs.~\eqref{eq:S4_B_SU_from_K} and use $\lambda_1+\gamma=r$, $\lambda_2+\gamma=-r/2$ and $\Omega^2=3r^2/4+4\omega^2$. This gives the other two moments:
\begin{align}
S(t)={}&S_\infty-\frac{2r}{\omega^2}k_1e^{\lambda_1t}+\frac{e^{\lambda_2t}}{\omega^2}\Big[(rk_2-2\Omega k_3)\cos(\Omega t)+(2\Omega k_2+rk_3)\sin(\Omega t)\Big],
\label{eq:S4_B_Ssol}\\
U(t)={}&U_\infty+\left(1+\frac{r^2}{2\omega^2}\right)k_1e^{\lambda_1t}+e^{\lambda_2t}\Bigg[-\left(1+\frac{r^2}{4\omega^2}\right)\big[k_2\cos(\Omega t)+k_3\sin(\Omega t)\big]+\frac{r\Omega}{2\omega^2}\big[k_2\sin(\Omega t)-k_3\cos(\Omega t)\big]\Bigg].
\label{eq:S4_B_Usol}
\end{align}
Thus the solution is controlled by one real mode and one damped oscillatory pair. At the threshold $\gamma=r$, the real exponent $\lambda_1$ vanishes and resonates with the constant source in Eq.~\eqref{eq:S4_B_Kode}, so that the second moments grow linearly in time.

\subsection{Integration constants}
The initial values of $K$, $S$ and $U$ fix the constants $k_i$:
\begin{subequations}\label{eq:S4_B_ICs}
\begin{align}
k_1+k_2&=K(0)-K_\infty,\\
-2rk_1+rk_2-2\Omega k_3&=\omega^2\left[S(0)-S_\infty\right],\\
\left(1+\frac{r^2}{2\omega^2}\right)k_1-\left(1+\frac{r^2}{4\omega^2}\right)k_2-\frac{r\Omega}{2\omega^2}k_3&=U(0)-U_\infty.
\end{align}
\end{subequations}
For an oscillator initially in the ground state, $K(0)=U(0)=\hbar\omega/4$ and $S(0)=0$. We solve Eqs.~\eqref{eq:S4_B_ICs} using $P(r)=0$ and Eqs.~\eqref{eq:S4_B_stationary}, and find
\begin{equation}
k_1=\frac{\hbar\omega\,\Gamma^{\rm HO}_B\,\omega^2}{2(r-\gamma)(3r^2+4\omega^2)},\qquad k_2=\frac{\hbar\omega}{4}-K_\infty-k_1,\qquad k_3=\frac{\omega^2S_\infty-2rk_1+rk_2}{2\Omega}.
\label{eq:S4_B_kground}
\end{equation}

\subsection{Mean occupation}
We add Eqs.~\eqref{eq:S4_B_Ksol} and \eqref{eq:S4_B_Usol} and use $\langle\hat n\rangle_t=E_t/\hbar\omega-1/2$:
\begin{equation}
\langle\hat n\rangle_t=\langle\hat n\rangle_\infty+\left(2+\frac{r^2}{2\omega^2}\right)\frac{k_1}{\hbar\omega}e^{\lambda_1t}+\frac{e^{\lambda_2t}}{\hbar\omega}\left[-\frac{r^2}{4\omega^2}\big(k_2\cos\Omega t+k_3\sin\Omega t\big)+\frac{r\Omega}{2\omega^2}\big(k_2\sin\Omega t-k_3\cos\Omega t\big)\right].
\label{eq:S4_B_nt}
\end{equation}
Here
\begin{equation}
\langle\hat n\rangle_\infty=\frac{K_\infty+U_\infty}{\hbar\omega}-\frac12=\frac{\Gamma^{\rm HO}_B(\gamma^2+4\omega^2)}{4\mathcal D_B},
\label{eq:S4_B_ninf}
\end{equation}
which is Eq.~\eqref{eq:BreuerStationaryNumber}. For the ground state, the amplitude of the real mode is
\begin{equation}
\left(2+\frac{r^2}{2\omega^2}\right)\frac{k_1}{\hbar\omega}=\frac{\Gamma^{\rm HO}_B(r^2+4\omega^2)}{4(r-\gamma)(3r^2+4\omega^2)}.
\end{equation}
Equation~\eqref{eq:S4_B_nt} satisfies $\langle\hat n\rangle_0=0$ and $\partial_t\langle\hat n\rangle|_0=\Gamma^{\rm HO}_B/4$, in agreement with Eq.~\eqref{eq:BreuerInitialHeating}. In the regime $\gamma,\Gamma^{\rm HO}_B\ll\omega$, one has $r\simeq\Gamma^{\rm HO}_B/2$, and the oscillating terms are suppressed by factors of order $r/\omega$. Equation~\eqref{eq:S4_B_nt} then reduces to
\begin{equation}
\langle\hat n\rangle_t\simeq\frac{\Gamma^{\rm HO}_B}{4(\gamma-\Gamma^{\rm HO}_B/2)}\left[1-e^{-(\gamma-\Gamma^{\rm HO}_B/2)t}\right],
\end{equation}
which is the solution of the secular Eq.~\eqref{eq:BreuerSecularNumber}. For $\gamma=0$ it gives $\langle\hat n\rangle_t\simeq[\exp(\Gamma^{\rm HO}_Bt/2)-1]/2$, consistently with Eq.~\eqref{eq:BreuerExponentialHeating}. Finally, the stationary moments entering Eq.~\eqref{eq:BreuerSpectrumMoments} follow from $\langle\hat x^2\rangle_{\rm ss}=2U_\infty/(m\omega^2)$ and $S_{\rm ss}=S_\infty$.

\section{Extension to composite systems}
\label{sec:compositeMain}

The results derived above refer to master equations for a single particle of mass $m$. Their application to composite systems, such as an optomechanical device, requires additional assumptions, and the replacement of $m$ by the total mass of the object is not automatic. Even for deterministic deformations, the inheritance of Planck-scale effects by composite bodies is model dependent. For deformed commutators, center-of-mass effects are suppressed by powers of the number of constituents in rigid motion but can be enhanced by internal motion~\cite{amelino-camelia_challenge_2013}. In Lie-algebra spacetimes, such as the one underlying the Arzano model, the composition law of momenta may itself be deformed. For the stochastic models considered here, one must in addition specify how the fluctuations act on, and are correlated between, the constituents.

Let us consider a body composed of $N$ identical constituents, each of mass $m$, so that its total mass is $M=Nm$. For definiteness, we identify these constituents with the atoms of the material. For a material containing several atomic species, replacing $m$ by the mean atomic mass gives provides a reasonable order-of-magnitude estimate. The choice of constituent is nevertheless part of the model. Identifying constituents with nucleons instead of atoms increases $N$ by a factor $A$, the mass number.

We describe the observed motion as a mode in which atom $a$, at equilibrium position $\mathbf r_a$, is displaced by $\mathbf u_a=\boldsymbol\phi_a q$, with $q$ the mode amplitude and $\boldsymbol\phi_a$ the dimensionless vector mode shape, normalized such that $\max_a|\boldsymbol\phi_a|=1$. The mode has effective mass $M_{\rm eff}=m\sum_a|\boldsymbol\phi_a|^2$ and momentum $\hat P=M_{\rm eff}\dot q$. Expanding the atomic momenta in normal modes, which for identical atoms obey $\sum_a\boldsymbol\phi_a^{\kappa}\cdot\boldsymbol\phi_a^{\kappa'}=0$ for $\kappa\neq\kappa'$, gives
\begin{equation}
\hat{\mathbf p}_a=\frac{m\boldsymbol\phi_a}{M_{\rm eff}}\hat P+\hat{\boldsymbol\pi}_a,
\label{eq:modeMomentum}
\end{equation}
where $\hat{\boldsymbol\pi}_a$ collects the contribution of all other modes and commutes with $\hat q$ and $\hat P$. We assume that the Hamiltonian separates into the mode of interest and the remaining degrees of freedom, $\hat H_N=\hat H_{\rm mode}+\hat H_{\rm rest}$, as for normal modes in the harmonic approximation, including those of clamped resonators, and denote the reduced state of the mode by $\hat\rho_{\rm mode}=\operatorname{Tr}_{\rm rest}\hat\rho_N$.

Rigid center-of-mass motion corresponds to a fixed unit vector $\boldsymbol\phi_a$, for which $M_{\rm eff}=M$. Treating the three directions together, the mode variables become the center-of-mass operators
\begin{equation}
\hat{\mathbf X}_{\mathrm{CM}}=\frac{1}{N}\sum_{a=1}^{N}\hat{\mathbf x}_a,\qquad \hat{\mathbf P}_{\mathrm{CM}}=\sum_{a=1}^{N}\hat{\mathbf p}_a,\qquad \hat K_{\mathrm{CM}}=\frac{|\hat{\mathbf P}_{\mathrm{CM}}|^2}{2M},
\end{equation}
and $\hat{\boldsymbol\pi}_a=\hat{\mathbf p}_a-\hat{\mathbf P}_{\rm CM}/N$ reduces to the internal momenta, which satisfy $\sum_a\hat{\boldsymbol\pi}_a=0$. In this case the separation of the Hamiltonian, $\hat H_N=\hat H_{\mathrm{CM}}+\hat H_{\mathrm{int}}$, holds for an isolated body or a body in a harmonic trap acting identically on all constituents, and we write $\hat\rho_{\mathrm{CM}}=\operatorname{Tr}_{\mathrm{int}}\hat\rho_N$. Neither case requires rigid motion. Rigidity enters only through the approximations discussed below, which neglect $\hat{\boldsymbol\pi}_a$.

A natural, although not unique, constituent-level extension of the single-particle dissipator is
\begin{equation}
\mathcal L_X^{(N)}\hat\rho_N=-\lambda_X\sum_{a,b=1}^{N}C_{ab}\,[\hat F_a^{(X)},[\hat F_b^{(X)},\hat\rho_N]],
\label{eq:generalCompositeDissipator}
\end{equation}
where
\begin{align}
\hat F_a^{(A)}&=\hat{\mathbf p}_a, & \lambda_A&=\frac{\tau_Ac^2}{2\hbar^2},\\
\hat F_a^{(B)}&=\hat K_a, & \lambda_B&=\frac{\tau_B}{2\hbar^2},\\
\hat F_a^{(I)}&=\hat K_a^2, & \lambda_I&=\frac{8a_P^2\tau_I}{\hbar^2},
\end{align}
with $\hat K_a=|\hat{\mathbf p}_a|^2/2m$ and $a_P=m\ell_P^2/\hbar^2$. 
For the Arzano model, the scalar product of the two vector operators is implied, which extends the one-dimensional model isotropically, and we assume the undeformed composition law of momenta. The appropriate composition rule for macroscopic bodies is model dependent, an issue related to the soccer-ball problem~\cite{amelino-camelia_challenge_2013}.
Equation~\eqref{eq:generalCompositeDissipator} follows from averaging the stochastic Hamiltonian $\hbar\sqrt{2\lambda_X}\sum_a\xi_a(t)\hat F_a^{(X)}$ over white noises with $\langle\xi_a(t)\xi_b(t')\rangle=C_{ab}\,\delta(t-t')$, with independent components $\langle\xi_{a,i}(t)\xi_{b,j}(t')\rangle=\delta_{ij}C_{ab}\,\delta(t-t')$ for the Arzano model, so that $C_{ab}$ is the correlation between the fluctuations acting on constituents $a$ and $b$. The coherent correction of the Petruzziello--Illuminati model, which originates from the mean value $\overline{\beta}$ common to all constituents, is generalized accordingly as $4a_P\overline{\beta}\sum_{a}\hat K_a^2$.

As a correlation matrix, $C_{ab}$ is real, symmetric and positive semidefinite. We normalize it as $C_{aa}=1$, so that each constituent taken individually obeys the single-particle master equation, and define
\begin{equation}
\mathcal C=\sum_{a,b=1}^{N}C_{ab},
\end{equation}
which ranges from $0$ to $N^2$. Two reference cases are
\begin{align}
C_{ab}=\delta_{ab}&\quad\Longrightarrow\quad\mathcal C=N,\label{eq:independentComposite}\\
C_{ab}=1&\quad\Longrightarrow\quad\mathcal C=N^2,\label{eq:collectiveComposite}
\end{align}
corresponding to independent and fully collective fluctuations, respectively. For statistically homogeneous fluctuations, $C_{ab}=C(\mathbf r_a-\mathbf r_b)$, with $C(\mathbf 0)=1$ and $C$ decaying over a correlation length $\ell_c$, for instance $C(\mathbf r)=e^{-r^2/2\ell_c^2}$ for Gaussian correlations. The independent and collective cases then correspond to $\ell_c$ much shorter than the interatomic distance and much longer than the size of the body, respectively. The same function describes the correlations with the atoms of the reference against which the displacement is measured, such as a support or another test mass, and enters through their relative displacement.

To obtain closed equations for the mode, we neglect the momenta $\hat{\boldsymbol\pi}_a$ of Eq.~\eqref{eq:modeMomentum} in the collapse operators, so that
\begin{equation}
\hat{\mathbf p}_a\simeq\frac{m\boldsymbol\phi_a}{M_{\rm eff}}\hat P,\qquad \hat K_a\simeq\frac{m|\boldsymbol\phi_a|^2}{M_{\rm eff}}\hat K,\qquad \hat K_a^2\simeq\frac{m^2|\boldsymbol\phi_a|^4}{M_{\rm eff}^2}\hat K^2,
\label{eq:rigidCompositeApproximation}
\end{equation}
with $\hat K=\hat P^2/2M_{\rm eff}$ the kinetic energy of the mode. For center-of-mass motion this is the rigid-motion approximation, $\hat{\mathbf p}_a\simeq\hat{\mathbf P}_{\rm CM}/N$, $\hat K_a\simeq\hat K_{\rm CM}/N$ and $\hat K_a^2\simeq\hat K_{\rm CM}^2/N^2$. The approximation does not assume that the momenta $\hat{\boldsymbol\pi}_a$ of a real solid vanish, which they do not even at zero temperature, but neglects their fluctuations and their coupling to the mode in deriving its reduced dynamics. 
Inserting Eq.~\eqref{eq:rigidCompositeApproximation} into Eq.~\eqref{eq:generalCompositeDissipator} gives the dissipators
\begin{subequations}
\label{eq:modeDissipators}
\begin{align}
\mathcal L_A\hat\rho_{\rm mode}&=-\lambda_A\frac{m_A}{M_{\rm eff}}[\hat P,[\hat P,\hat\rho_{\rm mode}]],\\
\mathcal L_B\hat\rho_{\rm mode}&=-\lambda_B\,\eta_B\,[\hat K,[\hat K,\hat\rho_{\rm mode}]],\\
\mathcal L_I\hat\rho_{\rm mode}&=-\lambda_I\,\eta_I\,[\hat K^2,[\hat K^2,\hat\rho_{\rm mode}]],
\end{align}
\end{subequations}
with
\begin{equation}
m_A=\frac{m^2}{M_{\rm eff}}\sum_{a,b}C_{ab}\,\boldsymbol\phi_a\cdot\boldsymbol\phi_b,\qquad \eta_B=\frac{m^2}{M_{\rm eff}^2}\sum_{a,b}C_{ab}|\boldsymbol\phi_a|^2|\boldsymbol\phi_b|^2,\qquad \eta_I=\frac{m^4}{M_{\rm eff}^4}\sum_{a,b}C_{ab}|\boldsymbol\phi_a|^4|\boldsymbol\phi_b|^4.
\end{equation}
These have the form of the single-particle dissipators for a particle of mass $M_{\rm eff}$, and give the oscillator rates
\begin{equation}
\Gamma_A^{\rm HO}=\frac{m_A\,\omega c^2\tau_A}{\hbar},\qquad \Gamma_B^{\rm HO}=\eta_B\,\tau_B\,\omega^2,\qquad \Gamma_I^{\rm HO}=8\eta_I\,a_P^2\tau_I\hbar^2\omega^4,
\label{eq:modeRates}
\end{equation}
with $\omega$ the mode frequency and $a_P$ the atomic value. 
Whether neglecting $\hat{\boldsymbol\pi}_a$ is justified depends on the model and on $C_{ab}$, as discussed below.

For the Arzano model, the first of Eqs.~\eqref{eq:modeDissipators} is exact for any mode, any $C_{ab}$ and any state, given Eq.~\eqref{eq:generalCompositeDissipator} and the separation of the Hamiltonian, and the approximation~\eqref{eq:rigidCompositeApproximation} is not required. Indeed, after inserting Eq.~\eqref{eq:modeMomentum} into the double commutator, every term containing $\hat{\boldsymbol\pi}_a$ is a commutator with an operator acting only on the other modes, and therefore vanishes under the partial trace. 
The mode thus behaves as a single particle of mass $M_{\rm eff}$, with the rate $\Gamma_A^{\rm HO}$ of Eq.~\eqref{eq:modeRates}.
For the center of mass, $m_A=M\mathcal C/N^2$, and
\begin{equation}
\partial_t\hat\rho_{\mathrm{CM}}=-\frac{i}{\hbar}[\hat H_{\mathrm{CM}},\hat\rho_{\mathrm{CM}}]-\tau_A\frac{c^2}{2\hbar^2}\frac{\mathcal C}{N^2}[\hat{\mathbf P}_{\mathrm{CM}},[\hat{\mathbf P}_{\mathrm{CM}},\hat\rho_{\mathrm{CM}}]].
\label{eq:ArzanoCompositeMain}
\end{equation}
For independent fluctuations, $m_A=m$ for any mode. For collective fluctuations, $m_A=m^2|\sum_a\boldsymbol\phi_a|^2/M_{\rm eff}$, which equals $M$ for the center of mass but vanishes for the internal modes of an isolated body, since these carry no net momentum. For $C_{ab}=C(\mathbf r_a-\mathbf r_b)$ with $\ell_c$ much larger than the lattice constant, the sum can be replaced by an integral, and a plane-wave mode of wave vector $\mathbf q$ gives $m_A\simeq\varrho\,\tilde C(\mathbf q)$, with $\varrho$ the mass density and $\tilde C(\mathbf q)=\int d^3r\,C(\mathbf r)e^{i\mathbf q\cdot\mathbf r}$. For $\ell_c\ll\lambda/2\pi$, with $\lambda$ the acoustic wavelength, $m_A\simeq\varrho\int d^3r\,C(\mathbf r)$ is the mass contained in a correlation volume, equal to $(2\pi)^{3/2}\varrho\ell_c^3$ for Gaussian correlations. For Gaussian correlations in the continuum plane-wave limit, $m_A$ reaches its maximum, $m_A\simeq0.6\,\varrho(\lambda/2)^3$, at $\ell_c\simeq0.3\lambda$, and decreases beyond.

For the Breuer model, neglecting the internal momenta, the center-of-mass equation is
\begin{equation}
\partial_t\hat\rho_{\mathrm{CM}}=-\frac{i}{\hbar}[\hat H_{\mathrm{CM}},\hat\rho_{\mathrm{CM}}]-\frac{\tau_B}{2\hbar^2}\frac{\mathcal C}{N^2}[\hat K_{\mathrm{CM}},[\hat K_{\mathrm{CM}},\hat\rho_{\mathrm{CM}}]].
\label{eq:BreuerCompositeMain}
\end{equation}
For a general mode, Eq.~\eqref{eq:modeMomentum} gives
\begin{equation}
\hat K_a=\frac{m|\boldsymbol\phi_a|^2}{M_{\rm eff}}\hat K+\frac{\hat P\,\boldsymbol\phi_a\cdot\hat{\boldsymbol\pi}_a}{M_{\rm eff}}+\frac{|\hat{\boldsymbol\pi}_a|^2}{2m}.
\label{eq:KaDecomposition}
\end{equation}
For fully collective fluctuations, the second term vanishes upon summation over $a$ by the orthogonality of the modes, so that $\sum_a\hat K_a=\hat K+\hat K_{\rm rest}$, with $\hat K_{\rm rest}$ the kinetic energy of the other modes, whose commutators vanish under the partial trace. The second of Eqs.~\eqref{eq:modeDissipators}, with $\eta_B=1$, is then exact.
For independent fluctuations, $\eta_B=\sum_a|\boldsymbol\phi_a|^4/(\sum_a|\boldsymbol\phi_a|^2)^2$ is the inverse number of atoms participating in the mode.
For a general $C_{ab}$, assuming that the state remains factorized, $\hat\rho_N\simeq\hat\rho_{\rm mode}\otimes\hat\rho_{\rm rest}$, with $\hat\rho_{\rm rest}$ stationary, the cross term generates the additional dissipator
\begin{equation}
-\frac{\tau_B W}{2\hbar^2}[\hat P,[\hat P,\hat\rho_{\rm mode}]],\qquad W=\sum_{a,b}C_{ab}\frac{\langle(\boldsymbol\phi_a\cdot\hat{\boldsymbol\pi}_a)(\boldsymbol\phi_b\cdot\hat{\boldsymbol\pi}_b)\rangle}{M_{\rm eff}^2},
\label{eq:BreuerCrossTerm}
\end{equation}
which vanishes for collective fluctuations, whereas the remaining terms vanish under the partial trace, because $\langle\hat{\boldsymbol\pi}_a\rangle=0$ and the momenta commute. In a homogeneous, isotropic solid, we define $k_BT_{\rm kin}\equiv\langle\hat\pi_{a,x}^2\rangle/m$, set by the thermal and zero-point motion of the atoms, so that $W=k_BT_{\rm kin}/M_{\rm eff}$ for independent fluctuations. Since the mean kinetic energy of every mode is at least $k_BT/2$, $T_{\rm kin}\geq T$, and the zero-point motion gives, in the Debye model, $T_{\rm kin}\geq3\Theta_D/8$, with $\Theta_D$ the Debye temperature. For silicon, $T_{\rm kin}\simeq240$~K even at millikelvin temperatures. Equation~\eqref{eq:BreuerCrossTerm} has the form of the Arzano dissipator with $c^2\tau_A\to W\tau_B$, and yields the state-independent heating rate $\Gamma_{B,\rm eff}/2$, with $\Gamma_{B,\rm eff}=\omega\tau_Bk_BT_{\rm kin}/\hbar$, independent of $N$ and $M_{\rm eff}$. It exceeds the direct contribution by $2k_BT_{\rm kin}/[\eta_B\hbar\omega(2n+1)]$, with $n$ the mean occupation of the mode, a factor of order $Nk_BT_{\rm kin}/\hbar\omega$ for the center of mass.

For the Petruzziello--Illuminati model, it is convenient to introduce $a_P^{\mathrm{CM}}=M\ell_P^2/\hbar^2=Na_P$. Neglecting the internal momenta, the center-of-mass equation reads
\begin{equation}
\partial_t\hat\rho_{\mathrm{CM}}=-\frac{i}{\hbar}\left[\hat H_{\mathrm{CM}}+4a_P^{\mathrm{CM}}\frac{\overline{\beta}}{N^2}\hat K_{\mathrm{CM}}^2,\hat\rho_{\mathrm{CM}}\right]-\tau_I\frac{8(a_P^{\mathrm{CM}})^2}{\hbar^2}\frac{\mathcal C}{N^6}[\hat K_{\mathrm{CM}}^2,[\hat K_{\mathrm{CM}}^2,\hat\rho_{\mathrm{CM}}]].
\label{eq:IlluminatiCompositeMain}
\end{equation}
Here internal momenta matter even for fully collective fluctuations, because $\sum_a\hat K_a^2$ does not separate into mode and rest operators. Squaring Eq.~\eqref{eq:KaDecomposition} gives
\begin{equation}
\hat K_a^2=\frac{m^2|\boldsymbol\phi_a|^4}{M_{\rm eff}^2}\hat K^2+\hat K\hat P R_a+\hat K G_a+\hat P h_a+\frac{|\hat{\boldsymbol\pi}_a|^4}{4m^2},
\label{eq:PICompleteExpansion}
\end{equation}
with
\begin{align}
R_a&=\frac{2m|\boldsymbol\phi_a|^2(\boldsymbol\phi_a\cdot\hat{\boldsymbol\pi}_a)}{M_{\rm eff}^2},\\
G_a&=\frac{|\boldsymbol\phi_a|^2|\hat{\boldsymbol\pi}_a|^2+2(\boldsymbol\phi_a\cdot\hat{\boldsymbol\pi}_a)^2}{M_{\rm eff}},\\
h_a&=\frac{(\boldsymbol\phi_a\cdot\hat{\boldsymbol\pi}_a)|\hat{\boldsymbol\pi}_a|^2}{mM_{\rm eff}}.
\end{align}
For a factorized state with stationary, isotropic Gaussian internal momenta, the $G_aG_b$ terms yield a Breuer-type contribution with $\tau_B\to16a_P^2\tau_I\sum_{a,b}C_{ab}\langle G_aG_b\rangle$. This equals $400\,a_P^2\tau_I(k_BT_{\rm kin})^2$ to leading order for collective center-of-mass fluctuations, and $752\,\eta_B\,a_P^2\tau_I(k_BT_{\rm kin})^2$ for independent fluctuations. These are individual contributions, not the complete reduced dissipator. The $R_ah_b$ terms, for example, contribute at the same order in $T_{\rm kin}$ for independent fluctuations.
The $h_ah_b$ terms produce an Arzano-type contribution, with $c^2\tau_A\to16a_P^2\tau_I\sum_{a,b}C_{ab}\langle h_ah_b\rangle$. For independent fluctuations and isotropic Gaussian atomic momentum marginals this gives
\begin{equation}
\Gamma_{I,\rm eff}^{\rm(ind)}=560\,a_P^2\tau_I\frac{\omega\,(k_BT_{\rm kin})^3}{\hbar}.
\label{eq:IlluminatiCrossTerm}
\end{equation}
For collective center-of-mass fluctuations, the exact constraint $\sum_a\hat{\boldsymbol\pi}_a=0$ changes this coefficient. A permutation-symmetric Gaussian internal state with $\langle\hat\pi_{a,i}^2\rangle=mk_BT_{\rm kin}$ gives $\Gamma_{I,\rm eff}^{\rm(coll)}=160\,a_P^2\tau_I\omega(k_BT_{\rm kin})^3/\hbar$ at leading order in $1/N$. The value for a general mode or correlation matrix depends on internal-momentum correlations.
As a check on the missing mixed terms, for an independent-noise center-of-mass mode the complete initial dissipative energy rate, under the same Gaussian assumption, is
\begin{equation}
\left.\dot E\right|_I=8a_P^2\tau_I\omega^2\left[35\Theta^3+\frac{154\Theta^2}{N}\langle\hat K\rangle+\frac{76\Theta}{N^2}\langle\hat K^2\rangle+\frac{8}{N^3}\langle\hat K^3\rangle\right],\quad \Theta=k_BT_{\rm kin}.
\label{eq:PICompositeInitialHeating}
\end{equation}
In the low-excitation regime, the leading $\Theta^3$ term is independent of $N$, while subleading terms depend on the mode state. The last term is the direct contribution of Eq.~\eqref{eq:IlluminatiCompositeMain}, which dominates only when the kinetic energy of the mode per atom exceeds about $k_BT_{\rm kin}$, far beyond the regime of mechanical oscillators. Similarly, the coherent correction adds $20a_P\overline{\beta}k_BT_{\rm kin}\hat K$ to the mode Hamiltonian, a temperature-dependent renormalization of its kinetic energy. As for the Breuer model, these results assume a factorized state with stationary internal motion.

When the internal momenta are neglected, which is exact for the Arzano model and for collective Breuer fluctuations, and for the Petruzziello--Illuminati model gives only the direct contribution, Eqs.~\eqref{eq:ArzanoCompositeMain}, \eqref{eq:BreuerCompositeMain} and \eqref{eq:IlluminatiCompositeMain} have the same form as the single-particle equations with $m$ replaced by $M$, after the replacements
\begin{equation}
\tau_A\to\tau_A\frac{\mathcal C}{N^2},\qquad \tau_B\to\tau_B\frac{\mathcal C}{N^2},\qquad \overline{\beta}\to\frac{\overline{\beta}}{N^2},\qquad \tau_I\to\tau_I\frac{\mathcal C}{N^6},
\label{eq:compositeParameterReplacements}
\end{equation}
where the last two refer to the total-mass parameter $a_P^{\mathrm{CM}}$. 
In all cases, the amplitude-damping term $\gamma\mathcal D[\hat a]$ can be added exactly as in the main text, and the characteristic oscillator rates of the center of mass are
\begin{equation}
\Gamma_{A,\mathrm{CM}}^{\mathrm{HO}}=\frac{M\omega c^2\tau_A}{\hbar}\frac{\mathcal C}{N^2},\qquad \Gamma_{B,\mathrm{CM}}^{\mathrm{HO}}=\tau_B\omega^2\frac{\mathcal C}{N^2},\qquad \Gamma_{I,\mathrm{CM}}^{\mathrm{HO}}=8(a_P^{\mathrm{CM}})^2\tau_I\hbar^2\omega^4\frac{\mathcal C}{N^6}.
\label{eq:compositeOscillatorRates}
\end{equation}

For the center of mass, replacing $m$ by $M$ is thus justified for the Arzano and Breuer models only for collective fluctuations, whereas independent fluctuations reduce both rates by a factor $N$. For the Petruzziello--Illuminati model, even collective fluctuations reduce the decoherence rate by $N^4$ and the coherent correction by $N^2$ relative to this replacement. Except for the Arzano model, however, these reductions hold only when the internal motion is negligible. For the Breuer model with non-collective fluctuations, and for the Petruzziello--Illuminati model, internal-motion terms can dominate in the low-excitation regime $\hbar\omega(2n+1)\ll k_BT_{\rm kin}$, with the precise condition depending on $C_{ab}$ and the mode shape. Their leading phonon-heating rates are not suppressed by $N$. The composite-body scaling is therefore not fixed by the single-particle master equation alone, but depends on the constituent-level extension, on $C_{ab}$ and, except for the Arzano model, on the internal motion of the body.

\end{document}